\documentclass[11pt]{article}
\usepackage{physics}
\usepackage{bm}
\usepackage[margin=1in]{geometry}        
\usepackage{amsmath,amssymb,amsfonts}    
\usepackage{graphicx}                    
\usepackage{caption,subcaption}          
\usepackage{float}                       
\usepackage{booktabs}                    
\usepackage{hyperref}                    
\usepackage{cite}                        
\usepackage{color}                       
\usepackage{tikz}                        
\usepackage{algorithm}                   
\usepackage[noend]{algpseudocode}        
\usepackage{authblk}                     
\usepackage{soul}                        
\usepackage{amsthm}                      
\usepackage{mathtools}
\usepackage{listings}                    
\usepackage{algorithm}
\usepackage{algpseudocode}
\usepackage{enumerate}
\usepackage{lineno}

\usepackage{setspace}
\newcommand{\triplespacing}{\setstretch {1.3}}
\triplespacing

\newcommand{\R}{\mathbb{R}}

\newcommand{\nodes}{\mathcal{V}}
\newcommand{\elts}{\mathcal{E}}

\newcommand{\msh}{T}

\newtheorem{definition}{Definition}
\newtheorem{proposition}{Proposition}
\newtheorem{remark}{Remark}

\title{PHASE: Pauli Hierarchical Assembly on Subdivided Elements\\
for Quantum-Compatible Operator Synthesis}
\author{
    {\Large Tillman Philo and Caglar Oskay\footnote{Corresponding author address: VU Station B\#351831, 2301 Vanderbilt Place, Nashville, TN 37235. Email: caglar.oskay@vanderbilt.edu}}
}
\date{}

\begin{document}

\maketitle

\vspace{-4em} 

\begin{center}
    {\Large     
        Department of Civil and Environmental Engineering\\
        Department of Mechanical Engineering \\
    Vanderbilt University \\
    Nashville, TN 37235 \\
    }
\end{center}

\begin{abstract}
Efficiently decomposing finite element stiffness matrices into the Pauli basis is challenging due to the exponential growth of Pauli strings with problem size. A naive Pauli expansion requires $\Theta(8^{\lceil \log_2 N \rceil})$ operations, where $N$ denotes the number of degrees of freedom, rendering direct decomposition infeasible for large systems. Existing approaches 
exploit algebraic sparsity or operator structure but do not incorporate the geometric organization intrinsic to finite element discretizations, and consequently exhibit poor scaling for stiffness matrices. To address this problem, we introduce PHASE, a hierarchical, geometry-aware Pauli decomposition algorithm that leverages recursive mesh partitioning to organize element contributions across multiple spatial scales. PHASE employs a hybrid strategy that combines full- and reduced-space Tensorized Pauli Decomposition with Fast Walsh-Hadamard Transform-based aggregation to assemble global Pauli coefficients efficiently. We show that this approach yields a dimension-dependent reduction in the exponential scaling exponent of Pauli assembly asymptotic complexity relative to existing methods, reducing the cost from $2^{2{\lceil \log_2 N \rceil}}$ to $2^{\gamma_d{\lceil \log_2 N \rceil}}$ with $\gamma_d < 2$ under standard mesh regularity and balanced partition assumptions. These results substantially improve the feasibility of quantum-compatible operator synthesis for large-scale finite element models.
\end{abstract}

\section{Introduction}

The finite element method (FEM) is an essential computational method for the numerical solution of boundary value problems in structural mechanics, heat transfer, fluid dynamics, and a broad range of other engineering and scientific disciplines~\cite{zienkiewicz2005basis, zienkiewicz2013solid, brenner2007mathematical, hughes2000linear}. A discretized mesh $T$ of the problem domain $\Omega$ is used to assemble a global stiffness matrix $K$ from sparse local element contributions $\hat{K}_e$, reducing the problem to a linear system $Ku=f$ whose solution yields the nodal degree of freedom values across the mesh. Problems of engineering interest (e.g. structural integrity assessments of civil infrastructure, aerospace component design, high fidelity multiscale material analysis, etc.) produce systems that require substantial classical computing resources. 

Classical direct solvers based on sparse Cholesky or multifrontal LU factorization are robust and widely used in practice, but fill-in during factorization causes the factor matrices to be substantially denser than the original system matrix~\cite{davis2006direct, george1994computer}, leading to superlinear complexity for the multi-dimensional problems. Preconditioned iterative solvers such as conjugate gradient with algebraic multigrid (AMG) preconditioning can approach $O(N)$ complexity per solve in favorable cases~\cite{saad2003iterative, trottenberg2001multigrid}, but convergence degrades with problem conditioning. The dominant wall-clock cost in large-scale parallel computations increasingly derives from memory bandwidth and communication overhead rather than floating-point throughput~\cite{shalf2020future}. Moore's law and the approaching physical limits of classical silicon scaling further compound this issue, making clear that further computational gains cannot rely on brute-force classical scaling~\cite{shalf2020future, ascac2010exascale}.

Quantum computing offers a different approach: by exploiting superposition and entanglement, quantum linear systems algorithms have demonstrated the potential for exponential improvement in scaling with problem size relative to classical solvers for sparse, well-conditioned systems~\cite{harrow2009quantum, childs2017quantum, bravoprieto2023variational}. This prospect has motivated a growing body of work on quantum-compatible FEM formulations~\cite{montanaro2016quantum, deiml2025realization, arora2025implementation, alkadri2025quantum, wang2026qafe, schade2024wave, Kouskiya:2026a}, of which demands a quantum-compatible representation of the stiffness operator as an upstream prerequisite. Constructing this representation efficiently for large, geometrically unstructured FEM systems remains a computational bottleneck for end-to-end quantum FEM pipelines. This manuscript is focused on addressing this problem.

\begin{figure}
    \centering
    \includegraphics[width=\textwidth]{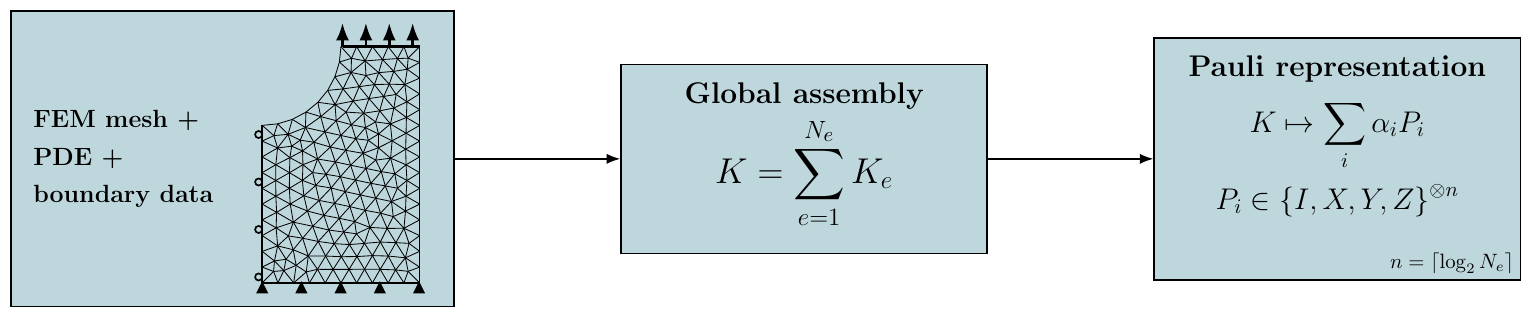}
    \caption{
        \emph{Quantum-compatible finite element assembly.}
        }
        \label{fig:problem-setting}
\end{figure}

Central to this challenge is the requirement that the system matrix $K$ can be encoded as a quantum circuit before any quantum solver can act on it. Since $K$ is non-unitary in general, direct circuit encoding is not possible, and a widely adopted strategy is the linear combination of unitaries (LCU) framework~\cite{childs2012hamiltonian, berry2015simulating}, which seeks to express $K$ as a weighted sum
\begin{equation}\label{eq:lcu}
A = \sum_i \alpha_i U_i,
\end{equation}
where each $U_i$ is a unitary operator and $\alpha_i \in \mathbb{C}$ is a known scalar weight. This decomposition enables the use of advanced techniques for quantum numerical methods, such as Variational Quantum Algorithms (VQAs)~\cite{cerezo2021variational,bravoprieto2023variational} and Hamiltonian simulation~\cite{berry2015simulating,lloyd1996universal}, that require the use of unitary-only operations. As such, efficient LCU representations of FEM stiffness matrices are essential for rendering quantum-compatible operator synthesis feasible in practice. Figure~\ref{fig:problem-setting} summarizes this problem setting.

Finding a suitable set of unitaries $\{U_i\}$ and corresponding weights
$\{\alpha_i\}$ satisfying the decomposition \eqref{eq:lcu} is nontrivial in
general. A widely used approach, particularly in quantum simulation and block-encoding schemes, is to expand operator $A$ in the Pauli basis~\cite{berry2015simulating,low2019hamiltonian,peruzzo2014variational,kandala2017hardware}:
\begin{equation} \label{eq:pauli-expansion}
    A = \sum_{P \in \mathcal P_n} \alpha_P P , \quad \alpha_P := \frac{1}{2^n}
\operatorname{Tr}(P A),
\end{equation}
where $\mathcal{P}_n$ is the complete set of
$4^n$ tensor product Pauli operators acting on $n$ qubits. The Pauli basis is a
natural and practical choice for several reasons. First, it forms an orthonormal
basis for the space of $2^n \times 2^n$ complex matrices under the Hilbert-Schmidt inner product, ensuring that any properly sized matrix admits a unique
expansion. Second, the coefficients $\alpha_P$ can be computed directly via
trace inner products, which allows for explicit evaluation.
Finally, each Pauli operator $P$ is itself unitary, enabling immediate
compatibility with LCU-based quantum algorithms, where unitarity of the summands
is essential.

Directly decomposing a $2^n \times 2^n$ operator into the Pauli basis requires
computing the weights for each Pauli string, resulting in a runtime complexity of $\Theta(8^n)$ if using dense 
matrix multiplication for computing $P A$. A range of techniques have
been proposed to mitigate this exponential scaling. The Tensorized Pauli
Decomposition (TPD) method of Hantzko et al. \cite{hantzko2024tensorized}
reduces the cost to $O(m^2 n 2^n)$ for matrices with $m$ nonzero entries, or
alternatively $O(mnp)$ and $O(2^n p^{3/2})$ under different sparsity
assumptions, where $p$ denotes the number of nonzero Pauli terms in the
expansion. However, for stiffness matrices arising from finite element
discretizations, both $m$ and $p$ scale with the global problem size $n$,
limiting the practical gains of TPD in this setting. Other approaches, such 
as the tree-based method of Koska et al.\cite{koska2024tree}, reduce constant 
factors via reuse of partial tensor products, while a Walsh-Hadamard-based scheme
from Georges et al. \cite{georges2025pauli} accelerates decomposition for structured cases such as
diagonal or banded matrices; neither approach generalizes effectively to sparse but
unstructured FEM matrices.

In this manuscript, we present the Pauli Hierarchical Assembly on Subdivided Elements (PHASE)
algorithm as a novel approach for representing finite element matrices in the Pauli basis. PHASE
is a hierarchical, structure-aware method that recursively partitions the mesh 
to construct a binary tree over the elements, enabling efficient aggregation
of local contributions at each depth of the hierarchy. The algorithm employs a hybrid
decomposition strategy: it applies either full-space TPD or reduced-space TPD depending
on an element's position in the partition tree, and at lower levels, PHASE uses the Fast
Walsh-Hadamard Transform (FWHT) to accelerate global aggregation of Pauli coefficients.
While TPD alone does not yield asymptotic improvements for FEM matrices due to their scaling structure, hybrid application of TPD within PHASE achieves an exponential
improvement in runtime relative to existing methods. 

Specifically, we propose an algorithm whose number of operations scales as $O(n2^{\gamma(\eta,d) n})$, where $\gamma(\eta, d)$, derived in Sections~\ref{sec:optimal-transition-depth}~and~\ref{sec:unbalanced-partition-analysis}, depends on the balancing quality of the recursive mesh partitioning and $d$ is the topological dimension of the mesh. For balanced partitions, 2D problems scale as $O(n2^{5n/3})$ and 3D problems as $O(n2^{7n/4})$ (Section~\ref{sec:optimal-transition-depth}). This achieves an exponential improvement over TPD, which for arbitrary FEM problems runs in $O(n2^{2n})$~\cite{hantzko2024tensorized}. In fact, as discussed in Section~\ref{sec:unbalanced-partition-analysis}, PHASE maintains this exponential improvement provided that no recursive partition produces a subdomain containing more than  approximately $71\%$ of the nodes of its parent.

The remainder of this work is outlined as follows. Section~\ref{sec:background} reviews the Pauli basis and the TPD algorithm~\cite{hantzko2024tensorized}. Section~\ref{sec:algorithmic-framework} establishes the geometric framework for PHASE, including recursive bisection construction, DoF encoding, and cut element scaling. Section~\ref{sec:hierarchical-pauli-assembly} described the PHASE assembly algorithm. Section~\ref{sec:asymptotic-runtime-model} derives the asymptotic runtime complexity under balanced and unbalanced partitions. Section~\ref{sec:numerical-validation} presents numerical validation. The conclusions are provided in Section~\ref{sec:conclusion}. Supporting derivations and supplementary concepts are included in Appendices~\ref{app:analytical-derivations}~and~\ref{app:supplementary-concepts}.


\section{Background}
\label{sec:background}

\subsection{Pauli Basis and Quantum-Compatible FEM}

Let $\mathcal{P}_n := \{I,X,Y,X\}^{\otimes n}$ denote the set of $n$-qubit Pauli strings. The collection $\mathcal{P}_n$ forms an orthonormal basis for the space $\mathbb{C}^{2^n \times 2^n}$ with respect to the Hilbert-Schmidt inner product
\begin{equation}
\langle A, B \rangle := \frac{1}{2^n} \operatorname{Tr}(A^\dagger B).
\end{equation}
Accordingly, any operator $A\in\mathbb{C}^{2^n \times 2^n}$ admits a unique Pauli expansion shown in Eq.~\ref{eq:pauli-expansion}. Each Pauli string $P$ is unitary, making such representations directly compatible with LCU constructions.

To apply this framework to finite element operators, consider a global stiffness matrix $K \in \mathbb{R}^{N \times N}$ arising from a conforming finite element discretization. Since $K$ is real and symmetric by construction, its Pauli expansion contains only real coefficients and is supported exclusively on Pauli strings with an even number of $Y$; however, PHASE does not explicitly exploit this restriction. Instead, the sparsity and locality structure that PHASE leverages is tied to the same physical origins that also force symmetry. Let
\begin{equation}
n:= \lceil \log_2 N \rceil,
\end{equation}
and define an embedding $\imath : \R^{N \times N} \hookrightarrow \mathbb{C}^{2^n \times 2^n}$ by zero-padding or index injection; the map $\imath$ acts purely as a scattering map, placing each entry of $K$ into its corresponding position in the larger matrix without mutation, with all remaining entries set to zero. Since $K$ is real symmetric, the image $\imath(K)$ lies entirely within the real symmetric subspace of $\mathbb{C}^{2^n \times 2^n}$. The codomain is taken to be $\mathbb{C}^{2^n \times 2^n}$ rather than $\mathbb{R}^{2^n \times 2^n}$ to maintain compatibility with the complex operator space of an $n$-qubit system, in which TPD and LCU-based methods are naturally defined. All Pauli decompositions in this work are performed on the embedded operator $\imath(K)$; for notational simplicity we suppress $\imath$ and write $K$ when no ambiguity arises. This embedding is purely algebraic and does not alter the sparsity or locality structure of the original operator. 

We say that an operator representation is \emph{quantum-compatible} if it satisfies the following properties:
\begin{enumerate}[i.]
    \item the operator is expressed as a linear combination of unitaries;
    \item the coefficients of this expansion are classically computable and explicitly accessible; and 
    \item the representation is suitable for block-encoding or related LCU-based quantum primitives.
\end{enumerate}
A Pauli expansion satisfies these requirements by construction.

FEM stiffness matrices possess strong geometric locality: they assemble as sums of element-level contributions and are sparse with respect to the nodal basis. However, they do not exhibit global tensor-product structure aligned with the Pauli basis. As a result, naive Pauli decomposition treats geometrically local interactions as globally dense operators in Pauli space, leading to exponential overhead even when the underlying FEM operator is sparse and structured. This structural mismatch motivates the need for decomposition strategies that respect the geometric organization of finite element operators while remaining compatible with Pauli-based quantum representations.

\subsection{Tensorized Pauli Decomposition (TPD)}
\label{sec:tpd-background}

The Tensorized Pauli Decomposition Algorithm (TPD) of Hantzko et al.~\cite{hantzko2024tensorized} provides an efficient method for computing the Pauli expansion of an operator $A \in \mathbb{C}^{2^n \times 2^n}$ by exploiting the tensor product structure of the Pauli basis. Rather than evaluating each coefficient via an explicit Frobenius inner product, TPD proceeds by recursively expanding the decomposition one tensor factor at a time. The key observation underlying TPD is that any $n$-qubit operator $A$ may be written as
\begin{equation}
A = \sum_{P\in \{I,X,Y,Z\}} P \otimes \Omega_{P*},
\end{equation}
where the matrices $\Omega_{P*} \in \mathbb{C}^{2^{n-1} \times 2^{n-1}}$, termed \emph{cumulative matrix weights}, collect the contributions of all Pauli strings whose first tensor factor is $P$. These matrices are computed by simple additions and subtractions of the four $(n-1)$-qubit blocks of $A$, and therefore require no matrix multiplication. Iterating this procedure on each nonzero $\Omega_{P*}$ yields a recursive factor-wise expansion that terminates after $n$ steps, at which point the Pauli coefficients $\{\alpha_i\}$ are recovered as scalars.

Algorithmically, TPD explores a quaternary tree over Pauli strings, pruning entire subtrees whenever a cumulative matrix weight vanishes. This allows the algorithm to automatically exploit tensor structure, sparsity, or symmetry in the input operator without prior knowledge of which Pauli terms are present. Both recursive and iterative variants of TPD are possible, as well as partial decompositions restricted to prescribed subsets of Pauli strings.

In the worst case, TPD computes all $4^n$ coefficients and attains a  complexity of $O(n4^n)$, matching the best known general bounds for Pauli decomposition. For structured inputs, significantly improved scaling is possible; for example, operators with strong tensor structure or few contributing Pauli strings may be decomposed in $O(mnp)$ or $O(2^n p^{3/2})$ time, where $m$ denotes the number of nonzero matrix entries and $p$ the number of nonzero Pauli terms~\cite{hantzko2024tensorized}. Importantly, these gains arise from algebraic structure in the operator itself rather than from reductions in the global qubit dimension.

For finite element stiffness matrices, however, both $m$ and $p$ scale with the total number of degrees of freedom, and TPD must operate on the full $2^n \times 2^n$ embedding of the operator. As a result, TPD alone does not yield asymptotic improvements for large, unstructured FEM systems. In PHASE, TPD is therefore treated as a local algebraic primitive and applied to smaller full- and reduced-space operators arising from element properties and hierarchical mesh partitioning, with global aggregation handled separately.

\subsection{Fast Walsh--Hadamard Transform (FWHT)}
\label{sec:fwht-intro}

The Walsh-Hadamard Transform is a linear transform on vectors indexed by binary strings, with coefficients weighted by parity of bitwise inner products. For a function $f:\{0,1\}^k \to \R$, it is defined by
\begin{equation}
(H^{(k)}f)(s) = \sum_{t\in \{0,1\}^k} (-1)^{\langle s, t \rangle} f(t),
\end{equation}
where $\langle s,t \rangle$ denotes the standard inner product over $\mathbb{Z}_2$.

The transform can be evaluated in $\Theta(k2^k)$ time via a recursive sequence of butterfly operations, yielding the Fast Walsh-Hadamard Transform (FWHT); a standard algorithmic description is given in Appendix~\ref{sec:fwht-algo}. Despite its name and common role in signal processing, in this work the FWHT is used not as a spectral analysis tool but as a structured summation operator. In particular, it provides an efficient mechanism for computing all parity-weighted sums over the Boolean cube $\{0,1\}^k$ simultaneously.

The utility of the FWHT in PHASE stems from this aggregation property. Many operator constructions described in sections below involve collections of quantities indexed by binary strings of fixed length, where each index corresponds to a distinct binary label. In such settings, naively forming all parity-weighted contributions requires explicit summation over $2^k$ terms per output coefficient, which is computationally prohibitive. FWHT replaces this explicit enumeration with a single structured transform whose cost scales as $\Theta(k2^k)$, independent of the number of distinct weighted sums being formed. Conceptually, it may be viewed as a change of basis on functions defined over $\{0,1\}^k$ that exposes all parity correlations at once.

The FWHT replaces this explicit enumeration with a single structured transform whose cost scales as $\Theta(k2^k)$, independent of the number of distinct weighted sums being formed. Conceptually, it may be viewed as a change of basis on functions defined over $\{0,1\}^k$ that exposes all parity correlations at once.

In PHASE, this capability is exploited to aggregate operator contributions that share a common reduced support but differ by a more refined binary sign sub-structure. The FWHT enables these contributions to be combined efficiently without flattening intermediate representations into the full Pauli basis. Its role is therefore purely algebraic and combinatorial: it acts as a global aggregation primitive that preserves tensor structure while avoiding exponential summation overhead.


\section{Algorithmic Framework of PHASE}
\label{sec:algorithmic-framework}

We begin by establishing notation for the computational mesh and its geometric structure.
This forms the foundation for the recursive domain decomposition introduced in
the following section.
\begin{definition}[Mesh] \label{def:mesh}
    Let $\Omega \subset \R^d$ be a bounded Lipschitz domain.
    A conforming, shape-regular mesh $\msh$ of $\Omega$ is
    a finite collection of bounded polytopes, with element set $\elts$
    and node set $\nodes$:
    \begin{equation}
        \msh = \{e\}_{e \in \elts},
        \qquad
        \nodes = \{v \in \overline{\Omega} : v \text{ is a node of some }
        e \in \elts\}.
    \end{equation}
    For each element $e \in \elts$, denote its node set by $\nodes(e) \subseteq \nodes$.
\end{definition}

\begin{definition}[Conformity, shape regularity, and quasi-uniformity]
    \label{def:conformity-etc}
    $\Omega$ for a global mesh size parameter $h > 0$ is a \emph{conforming, shape-regular, quasi-uniform} mesh $\msh = \{e\}_{e \in \elts}$ of bounded polytopes satisfying
    \begin{enumerate}
        \item \textbf{Conformity:} For any $e,e' \in \elts$ with $e \neq e'$, the 
        intersection $\overline{e} \cap \overline{e'}$ is either empty, a common
        vertex, edge, or face.
        \item \textbf{Shape regularity:} There exists a constant $C_\mathrm{shape}>0$
        independent of $h$ such that
        \begin{equation} \frac{h_e}{\rho_e} \leq C_{\mathrm{shape}} \qquad \forall e\in \elts, \end{equation}
        where $h_e$ is the diameter of $e$ and $\rho_e$ the radius of its inscribed
        sphere.
        \item \textbf{Quasi-uniformity:} There exist constants $c_1,c_2 > 0$ such that 
        \begin{equation} c_1 h \leq h_e \leq c_2 h, \qquad \forall e \in \elts. \end{equation}
    \end{enumerate}
\end{definition}

Under standard assumptions, the total number of elements and nodes scale linearly, so we use $N = |\mathcal{E}| \asymp |\mathcal{V}|$ to denote the characteristic size of the mesh, where $a \asymp b$ denotes order-equivalence: there exist constants $c_1, c_2 > 0$ such that $c_1 a \leq b \leq c_2 a$, a strictly weaker statement than $a/b \to 1$.

Each element $e \in \elts$ has volume $|e| \asymp h^d$, hence the global relation
\begin{equation}
    N \asymp \frac{|\Omega|}{h^d} \quad \Longrightarrow \quad h \asymp N^{-1/d}.
\end{equation}
This relation connects the mesh resolution $h$ with the number of elements $N$ and
is used below to characterize the asymptotic scaling of subsets of elements
within the mesh.

\subsection{Recursive Mesh Partitioning and Definition of Cut Elements}
\label{sec:recursive-mesh-partitioning}

The geometric structure of $\msh$ provides the natural foundation for recursive spatial subdivision. A bisection introduces a separator that simultaneously partitions the domain $\Omega$ into two subdomains and the mesh $\msh$ into two submeshes of approximately equal numbers of nodes, inducing a set of cut elements along the separator interface. Iterating this construction generates a binary hierarchy of nested subdomains and submeshes,
\begin{equation}
(\Omega, \msh) \;\to\; 
(\Omega_0, \msh_0) \sqcup (\Omega_1, \msh_1) \;\to\; 
(\Omega_{00}, \msh_{00}) \sqcup \cdots \sqcup (\Omega_{11}, \msh_{11}) 
\;\to\; \dots,
\end{equation}
whose levels serve as the aggregation levels in the PHASE assembly algorithm. Efficient algorithms for generating these separators in a geometry-respecting manner are well-established~\cite{miller1998geometric,lipton1979separator,alan1973nested}. 

We note that each node of the mesh is generally associated with one or more finite element degrees of freedom (DoF), depending on the problem. For instance, a scalar diffusion problem has one DoF per node, while an elasticity problem has multiple DoFs per node. When the global stiffness operator is expressed in the Pauli basis, these degrees of freedom map naturally to qubit indices, and the resulting hierarchical domain structure provides a direct geometric counterpart to the bitstring labeling described in Section~\ref{recursive-application-and-dof-encoding}. We now formalize this construction.

\begin{definition}[Separator] \label{def:separator}
    A \emph{separator} is a $(d-1)$-dimensional Lipschitz manifold (a level set)
    $\Sigma \subset \overline{\Omega}$ for which there exists a 
    Lipschitz function $s: \overline{\Omega} \rightarrow \R$ satisfying
    \begin{equation}
        \Sigma = \{x \in \overline{\Omega} \, | \, s(x)=0\}, \qquad
        \nabla s(x) \neq 0 \,\, \textrm{for} \,\, \mathcal{H}^{d-1} \,\, \textrm{-a.e.} \,\, x \in \Sigma.
    \end{equation}
    where $\mathcal{H}^{d-1}$ denotes the $(d-1)$-dimensional Hausdorff measure on $\Sigma$. The separator divides $\Omega$ into two open subdomains
    \begin{equation}
        \Omega_1 = \{x : s(x) > 0\}, \qquad \Omega_0 = \{x: s(x) < 0\},
    \end{equation}
    with unit normal vector field $n_{\Sigma}(x) = \nabla s(x)/ ||\nabla s(x)||$
    oriented from $\Omega_0$ to $\Omega_1$.
\end{definition}

To translate the geometric bisection of $\Omega$ into a combinatorial partition of the mesh, we assign each node $v \in \nodes$ a binary label $\ell(v) \in \{0,1\}$ determined by its position relative to $\Sigma$, where $\ell(v)=1$ indicates membership in the positive subdomain $\Omega_1$ and $\ell(v)=0$ indicates membership in the negative subdomain $\Omega_0$. This labeling is derived directly from the separator function $s(x)$ and is used in Definition~\ref{def:submeshes-and-cut-elts} to classify elements as belonging to $\msh_0$, $\msh_1$, or the cut set $\msh^\times$:
\begin{equation} \label{node_labeling}
    \ell : \nodes \to \{0,1\},
    \qquad
    \ell(v) =
    \begin{cases}
        1 \textrm{ if } s(v) \geq 0, \\
        0 \textrm{ if } s(v) < 0.
    \end{cases}
\end{equation}
Nodes that lie exactly on the separator are assigned to the $1$ side by convention.
A geometric justification of this convention is provided in 
Appendix~\ref{sec:geo-implementation-notes}.

\begin{definition}[Submeshes and cut elements] \label{def:submeshes-and-cut-elts}
    Using the node labels, we define the element subsets
    \begin{align}
        \elts_1 =& \{e \in \elts \,\, | \,\, \ell(v) = 1 \,\, \forall v \in \nodes(e)\}, \\
        \elts_0 =& \{e \in \elts \,\, | \,\, \ell(v) = 0 \,\, \forall v \in \nodes(e)\},
    \end{align}
    and the set of \emph{cut elements} are defined as
    \begin{align}
        \elts^\times = \{e \in \elts \,\, | \,\, \exists \, v,w \in \nodes(e) \textrm{ such that } \ell(v) \neq \ell(w)\}.
    \end{align}
    The set of cut elements includes all elements that the separator splits with the caveat of the assignment indicated in Eq.~\ref{node_labeling} when a node lies on the separator. The associated submeshes are
    \begin{align}
        \msh_{b} = \{e : e \in \elts_{b}\}, \qquad \msh^\times = \{e : e \in \elts^\times\}, \qquad b \in \{0,1\}.
    \end{align}
    We refer to $\msh_b$ as the submeshes and $\msh^\times$ as the \emph{cut set}.
\end{definition}

\begin{proposition}[Partition of the mesh]
    The mesh elements decompose uniquely as
    \begin{align}
        \elts = \elts_1 \sqcup \elts_0 \sqcup \elts^\times,
    \end{align}
    that is, each element lies either entirely in the $1$ submesh, 
    entirely in the $0$ submesh, or in the cut set.
\end{proposition}
The decomposition follows immediately from the disjointness of the defining conditions
on node labels and the discrete nature of the mesh.

\begin{figure}
    \centering
    \includegraphics[width=\textwidth]{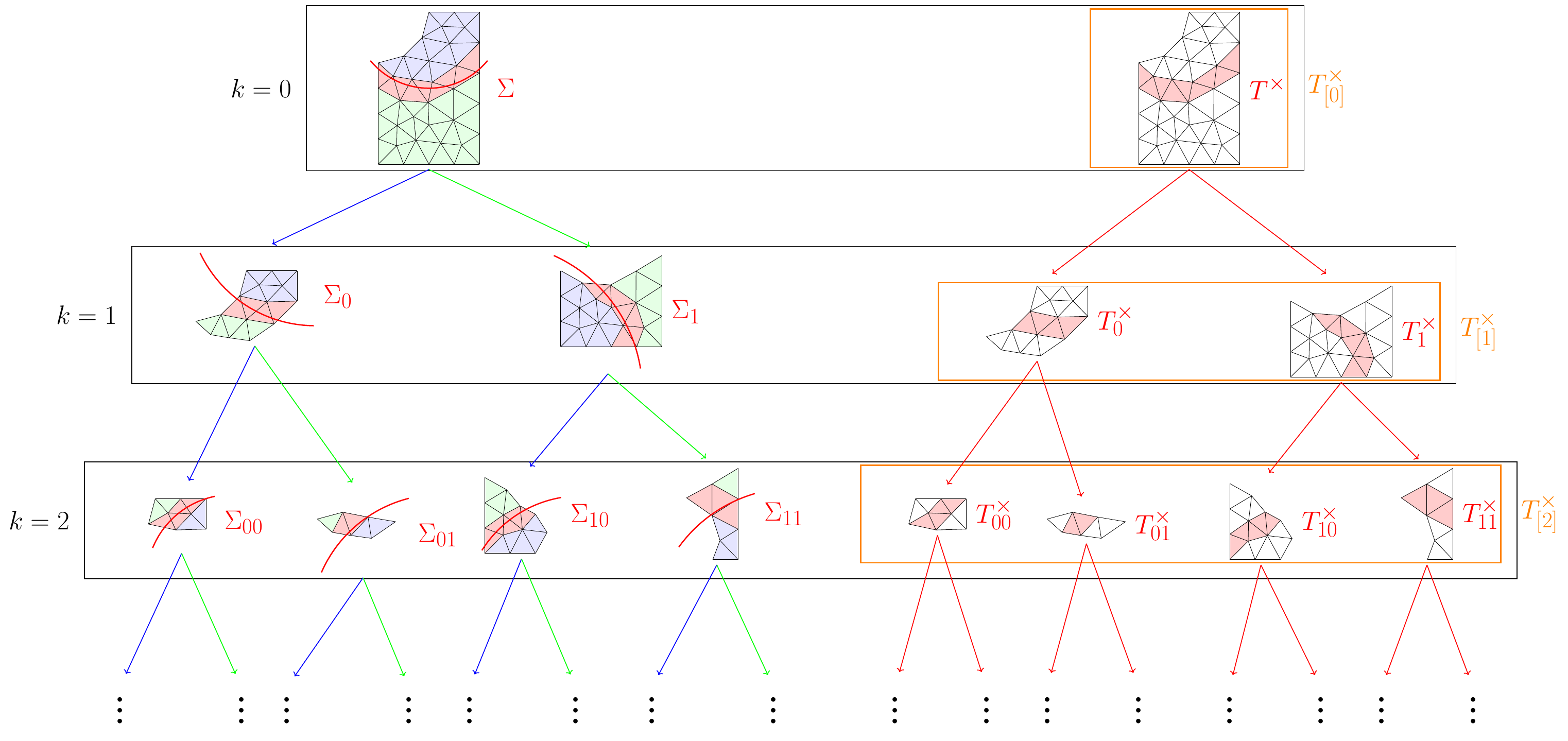}
    \caption{
        \emph{Recursive separator hierarchy and associated cut sets.}
    }
    \label{fig:partition-tree}
\end{figure}

This single-separator decomposition defines the geometric primitives used in the recursive hierarchy. Applying the same construction to each subdomain $\Omega,\,\Omega_0,\,\Omega_1,\,\dots$ at successive depths produces a family of separators $\Sigma,\,\Sigma_0,\,\Sigma_1,\,\dots$ and corresponding cut sets $T^\times,\,T^\times_{0},\,T^\times_{1},\,\dots,$, which together form the recursive partition tree underlying the PHASE algorithm. Figure~\ref{fig:partition-tree} shows a graphical illustration of the partitioning scheme.

\subsection{Recursive Application of Separators and DoF Encoding}
\label{recursive-application-and-dof-encoding}

We now introduce the recursive application of separators. Let $\Sigma$ denote the separator applied to the initial domain $\Omega$. The bisection produces two disjoint open subdomains
\begin{equation}
    \Omega_1 = \{x \in \Omega: s(x) \geq 0\}, \qquad 
    \Omega_0 = \{x \in \Omega : s(x) < 0\},
\end{equation}
together with the corresponding submeshes $T_1$, $T_0$, 
$T^\times$. We assign the separator at this level the \emph{depth index}
$k=0$.

For every nonempty subdomain $\Omega_{\bar{q}}$ produced at depth $k$, where
\begin{equation}
    \bar{q} = q_1 q_2 \dots q_k, \qquad q_i \in \{0,1\},
\end{equation}
denotes the \emph{sign sequence} (or \emph{branch label}) accumulated along the path
from the root, we define a new separator $\Sigma_{\bar{q}}$ by a local 
Lipschitz function
\begin{equation}
    s_{\bar{q}} : \overline{\Omega_{\bar{q}}} \to \R,
    \qquad 
    \Sigma_{\bar{q}} = \{x: s_{\bar{q}}(x)=0 \}.
\end{equation}
This induces the next pair of subdomains
\begin{equation}
    \Omega_{\bar{q}1} = 
    \{ x \in \Omega_{\bar{q}} : s_{\bar{q}} \geq 0 \},
    \qquad
    \Omega_{\bar{q}0} = 
    \{ x \in \Omega_{\bar{q}} : s_{\bar{q}} < 0 \}
\end{equation}
and the associated submeshes $T_{\bar{q}1}$, 
$T_{\bar{q}0}$, and $T_{\bar{q}}^\times$.
The recursion continues until each leaf domain $\Omega_{\bar{q}}$ contains
exactly one element or the number of elements is less than a predetermined threshold.

\begin{figure}
    \centering
    \includegraphics[width=0.5\textwidth]{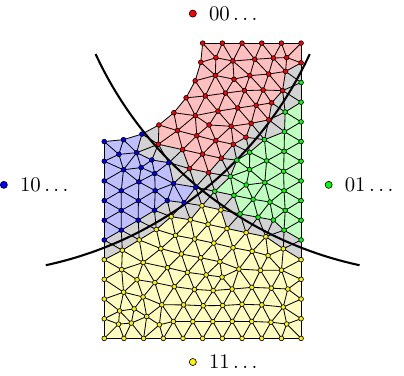}
    \caption{
        \emph{Illustration of recursive geometric partitioning and binary prefix labeling.}
    }
    \label{fig:structure-demo}
\end{figure}

Each full sign sequence $\bar{q}=q_1q_2\dots q_k$ uniquely identifies
a path from the root to a leaf in the binary partition tree. The family of all such
sequences at depth $k$ is $\{0,1\}^k$. Figure~\ref{fig:structure-demo} illustrates this correspondence at depth $k=2$.

The binary sign sequences $\bar{q}$ accumulated along each path of the partition tree serve a dual purpose: they record geometric location within the recursive hierarchy and provide the foundation for a canonical binary labeling of all degrees of freedom. We label the binary prefix $\bar{q}$ as a \emph{coarse address} at depth $k$ and its length ($k$) as its \emph{depth index}. $\bar{q}$ records the location of a 
subdomain in the recursive hierarchy and is used to enforce operator quantum compatibility. At the finest depth, where each subdomain $\Omega_{\bar{q}}$ contains a single element (or a predetermined number of elements),
all degrees-of-freedom (DoF) within that element share the same coarse address. To distinguish them, we introduce a \emph{local DoF label}.

Let each element $e$ have $\nu_e$ local degrees of freedom, which in most standard problems is a uniform constant $\nu_e = \nu$. Define $\nu_{\max} = \max_{e \in \elts} \nu_e$. Each local DoF is assigned a universal $\lceil\log_2\nu_{\max}\rceil$-bit code
\begin{equation}
\eta(i) \in \{0,1\}^{\lceil \log_2 \nu_{\max} \rceil}, \qquad i = 1, \dots, \nu_{\max}
\end{equation}
using a consistent scheme that applies to every element. The local code $\eta(i)$ is used to distinguish DoFs within an element once the coarse address is fixed.

Hence each global DoF corresponds to a concatenated bitstring
\begin{equation}
    b = \bar{q} \;\|\; \eta(i) \in \{0,1\}^{n}
\end{equation}
which defines a one-to-one mapping between mesh DoFs and computational basis states in
an $n$-qubit Hilbert space, where $n=k_{\max} + \lceil\log_2\nu_{\max}\rceil$ and $k_{\textrm{max}}$ is the maximum separator recursion depth. This distinction between $\bar{q}$
(geometric hierarchy) and $\eta$ (local structure) ensures that DoFs associated with
the same node but different components or basis functions are uniquely defined.

\subsection*{Treatment of unbalanced partitions}

The preceding construction assumes ideal bisections of equal size, but in practice
separators do not divide the mesh exactly in half. Some branches of the recursive tree
may terminate earlier than others, producing coarse addresses $\bar{q}$ of varying length. Since the global bitstring encoding requires all DoF addresses to have uniform length $n = k_{\max} + \lceil \log_2 \nu_{\max} \rceil$, shorter addresses must be extended to match. Let $k(\bar{q})$ denote the depth at which the branch labeled by $\bar{q}$ terminates, and define the maximum depth $k_{\max} = \max_{\bar{q}} k(\bar{q})$.

To reconcile this length mismatch we employ \emph{virtual completion}: for any branch that terminates at depth $k < k_{\max}$, its coarse address $\bar{q}$ is extended to length $k_{\max}$ by padding zeros,
\begin{equation}
    C(\bar{q}) = \bar{q} \;\|\; 0^{k_{\max} - k}
\end{equation}
The extension is \emph{virtual} in the sense that the appended zeros do not correspond to any real geometric subdivision; no further bisection of the leaf domain is performed. This is a purely algebraic operation that preserves the hierarchical structure while ensuring that all addresses have equal length, allowing unbalanced trees to be embedded into the global bit labeling scheme without forcing premature termination of other branches. Each leaf domain is then uniquely identified by the completed address $C(\bar{q})$, and every DoF in the mesh is labeled by
\begin{equation}
    b = C(\bar{q}) \;\|\; \eta(i).
\end{equation}

The completed binary representation provides a uniform encoding of every active
DoF across the mesh. After virtual completion, each leaf subdomain is identified
by its padded coarse address $C(\bar{q}) \in \{0, 1\}^{k_{\max}}$, and each local 
DoF within that subdomain is distinguished by its local label $\eta(i) \in \{0,1\}^{\nu_{\textrm{max}}}$.

Concatenating these strings defines the global bitstring
\begin{equation}
    b = C(\bar{q}) \;\|\; \eta(i) \in \{0,1\}^n, \quad n = k_{\max} + \lceil\log_2\nu_{\textrm{max}}\rceil
\end{equation}
The set of all valid bitstrings,
\begin{equation}
    \mathcal{B} = \{C(\bar{q}) \;\|\; \eta(i) \,|\, \Omega_{\bar{q}}
    \textrm{ is a leaf subdomain and } i \textrm{ indexes an active DoF}\},
\end{equation}
constitutes the complete binary address space of the mesh.

Each element of $\mathcal{B}$ corresponds uniquely to a computational basis state in the 
$n$-qubit Hilbert space,
\begin{equation}
    b \longleftrightarrow |b \rangle = |C(\bar{q}) \rangle \otimes |\eta(i) \rangle.
\end{equation}
This mapping provides the canonical identification between the finite element DoFs and quantum basis vectors. It compactly encodes both geometric hierarchy (through $C(\bar{q})$) and intra-element structure (through $\eta(i)$) within a uniform register length, even when the underlying mesh partitions are unbalanced. In what follows, this labeling is used to define tensor-product projectors that isolate or aggregate contributions from specific subdomains in the global Pauli assembly.

\subsection{Asymptotic Scaling of Cut Elements}
\label{sec:asymptotic-scaling-of-cut-elts}

We estimate the number of mesh elements intersected by a separator surface, i.e., the number of \emph{cut elements}. This estimate is employed in the asymptotic complexity analysis of PHASE. Let $d$ denote the Hausdorff (topological) dimension of $\msh$ as defined in Definition~\ref{def:mesh}, and let $N = |\mathcal{E}|$ denote the total number of elements. Here $d$ refers to the intrinsic dimension of the mesh, not that of the embedding space; under the Lipschitz regularity assumed throughout, the Hausdorff dimension coincides with the standard topological dimension and takes integer values only. The scaling of the number of cut elements  ($\#\text{ cut elements} \asymp N^{(d-1)/d}$) depends solely on how mesh volume and separator area scale with element size, and is unchanged if $\msh$ is embedded in a higher-dimensional space without refinement in the additional dimensions.

Under the standard assumptions of Section~\ref{sec:algorithmic-framework}, the characteristic mesh size and element count satisfy $h \asymp N^{-1/d}$. We will use this to understand the asymptotic characteristics of the number of cut elements at a given depth.

Let $\Sigma$ denote a separator surface with finite, mesh-independent $(d-1)$-dimensional measure $|\Sigma| = O(1)$. An element is classified as a cut element if its support intersects $\Sigma$. For a quasi-uniform mesh with characteristic element diameter $h$, any such element intersects $\Sigma$ in a region whose $(d-1)$-dimensional measure is $O(h^{d-1})$. Since these intersection patches are disjoint up to constants depending only on shape regularity, the number of cut elements satisfies
\begin{equation}
    \#\{\textrm{cut elements}\} \asymp \frac{|\Sigma|}{h^{d-1}} \asymp h^{-(d-1)}.
\end{equation}
Substituting $h \asymp N^{-1/d}$ yields the scaling law
\begin{equation}\label{eq:no-cut-elts}
    \#\{\textrm{cut elements}\} \asymp N^{(d-1)/d}.
\end{equation}
This relation shows that the number of cut elements grows \emph{sublinearly}
with the total number of elements $N$. The fraction of cut elements therefore 
decreases as the mesh is refined---a structural property that underpins the efficiency
of a hierarchical and multilevel assembly algorithm such as PHASE. 

If we take into account the exponential scaling of submeshes relative to partition depth, we can characterize the number of total cut elements at a given depth across all submeshes, which is denoted by $|T_{[k]}^\times|$. The result is
\begin{equation}
|T_{[k]}^\times|\asymp 2^{\beta_d n + \beta_d^- k}
\end{equation}
where
\begin{equation}
    \beta_d := \frac{d-1}{d},
    \qquad
    \beta_d^- := 1 - \beta_d = \frac{1}{d},
    \qquad
    \beta_d^+ := 1 + \beta_d = \frac{2d-1}{d}.
\end{equation}
We have introduced $\beta_d^+$ in addition to $\beta_d$ and $\beta_d^-$ seen in the asymptotic scaling for $|T_{[k]}^\times|$ as $\beta_d^+$ arises in the complexity analyses discussed below and is explicitly related to $\beta_d$ and $\beta_d^-$. The exponent $\beta_d$ characterizes the asymptotic growth rate of the number of cut elements with respect to $N$, while the conjugate quantities $\beta_d^\pm$ serve as shorthand for dimension-dependent exponents that arise during analytical derivations associated with the asymptotic runtime complexity model of the algorithm. A detailed geometric derivation of this scaling law is provided in Appendix~\ref{app:derivation-of-cut-elt-scaling}.


\section{Hierarchical Pauli Assembly}
\label{sec:hierarchical-pauli-assembly}

\subsection{Global Stiffness Assembly in the Pauli Basis}

We seek to assemble the global stiffness operator $K \in \R^{2^n \times 2^n}$
explicitly in the $n$-qubit Pauli basis $\{I,X,Y,Z\}^{\otimes n}$, in a way that exploits
the hierarchical structure created by the coordinate-wise separator labeling scheme (Section~\ref{recursive-application-and-dof-encoding}). We begin by setting notation for the local-to-global embedding of element stiffness operators. The \emph{local} element stiffness operator for an element $e \in \elts$ is denoted as $\hat{K}_e \in \R^{\nu_e \times \nu_e}$. Let $L_e \in \{0, 1\}^{2^n \times \nu_e}$ be the (boolean) assembly matrix that injects element
DoFs into the global DoF space (its columns contain ``1'' at the global row associated with each local DoF and zeros elsewhere). The \emph{global embedding} of the local element matrix is then
\begin{equation}
    K_e = L_e \hat{K}_e L_e^{\top} \in \R^{2^n \times 2^n}.
\end{equation}
This convention explicitly distinguishes local element operators from global element operators. With these definitions, the classical finite-element stiffness operator is
\begin{equation}
    K = \sum_{e \in \elts} K_e = \sum_{e \in \elts} L_e \hat{K}_e L_e^\top.
\end{equation}
Any real symmetric $K$ admits a unique Pauli expansion according to Eq.~\eqref{eq:pauli-expansion}. We use $\mathcal{P}(\cdot)$ to denote the Pauli representation of an operator in its appropriate Hilbert space size:
\begin{equation}
    \mathcal{P}(K) = \{(\alpha_P, P) \,|\, P \in \{I,X,Y,Z\}^{\otimes n}, \,\alpha_P\neq 0\}.
\end{equation}

A key consequence of the recursive partitioning outlined in Section~\ref{sec:recursive-mesh-partitioning} is that every element can be assigned to a unique level of the hierarchy, enabling a level-wise decomposition of the global Pauli assembly. Let $\Sigma_{[k]} := \{\Sigma_{\bar{q}} \, | \, {|\bar{q}|=k}\}$ denote the set of separators 
generated by the recursive bisection of $\Omega$ at depth $k$. Similarly, $T_{[k]}^\times := \{T_{\bar{q}}^\times \,|\, |\bar{q}|=k\}$ contains all elements that were cut by a depth $k$ separator (cut by an element of $\Sigma_{[k]}$). In practice, recursion terminates when a subgraph falls below a prescribed size threshold rather than every element has been individually cut. To ensure that every element belongs to exactly one node of the partition tree, all elements remaining in a subdomain $\Omega_{\bar{q}}$ at termination are assigned collectively to a leaf in the cut element partition tree following its recursive labeling path. Formally, we extend the definition of $\msh_{\bar{q}}^\times$ to include all uncut elements within $\Omega_{\bar{q}}$. under this convention, every element contribute exactly once at the depth it was cut or assigned. 

Before stating the global identity, we define a merge operation on Pauli coefficient lists. For two such lists $\mathcal{A} = \{(\alpha_P, P)\}$ and $\mathcal{B} = \{(\beta_P, P)\}$, define
\begin{equation}
    \mathcal{A} \uplus \mathcal{B} = \{(\alpha_P + \beta_P,\, P) \mid P \in \mathrm{supp}(\mathcal{A}) \cup \mathrm{supp}(\mathcal{B}),\ \alpha_P + \beta_P \neq 0\}
\end{equation}
where $\alpha_P=0$ for $P \notin \mathrm{supp}(\mathcal{A})$ and $P \notin \mathrm{supp}(\mathcal{B})$. The operation $\uplus$ is associative and commutative, so iterated merges are well-defined regardless of order.

Consequently, the global Pauli expansion can be written as a level-wise hierarchical merge
\begin{equation}
    \mathcal{P}(K) = \biguplus_{k=0}^{k_{\max}}\biguplus_{e \in T^{\times}_{[k]}} \mathcal{P}(K_e)
\end{equation}
This identity is simply a re-indexing of $\biguplus_{e \in \elts} \mathcal{P}(K_e)$ by the first
cut where an element appears and is the key structural relation exploited by PHASE's hybrid
assembly. It isolates at depth $k$, a geometrically thin set of cut elements whose cardinality
scales sublinearly with the global problem size (cf. the $N^{(d-1)/d}$ scaling of cut sets),
and it preserves exactness: no element is double-counted, and no contribution is omitted.

\begin{figure}
    \centering
    \includegraphics[width=1.0\textwidth]{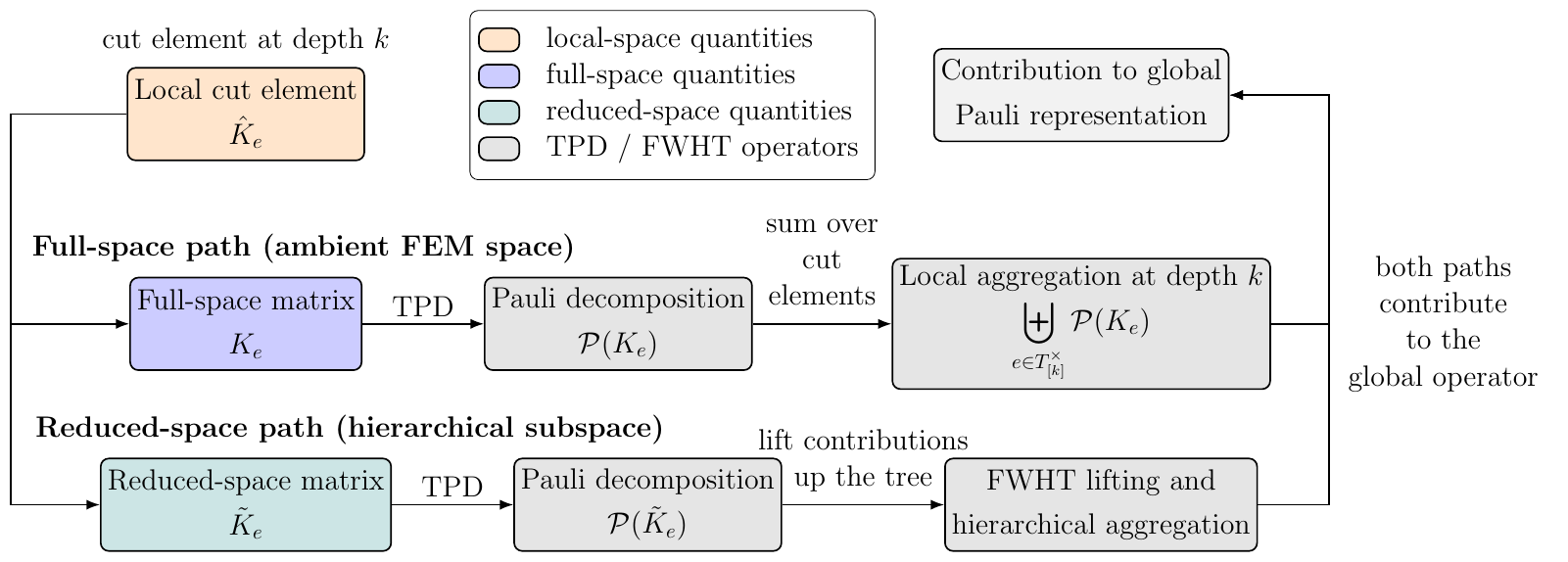}
    \caption{
        \emph{Dual assembly paths for hierarchical pauli decomposition of cut elements.}
    }
    \label{fig:branch-diagram}
\end{figure}

The PHASE assembly scheme processes cut elements at each depth of the recursive hierarchy through two complementary paths, illustrated in Figure~\ref{fig:branch-diagram}. In the full-space path, local stiffness contributions are embedded directly in the global Hilbert space and decomposed via TPD. In the reduced-space, contributions are first decomposed in a local subspace and then lifted to global space via FWHT-based aggregation. The following subsections develop each component of this scheme in detail.

\subsection{Full-Space Tensorized Pauli Decomposition (TPD)}
\label{sec:full-space-tpd}

At each depth $k$ of the recursive hierarchy, the PHASE algorithm processes the corresponding set of cut elements $T_{[k]}^\times$. The first assembly options applies TPD (Section \ref{sec:tpd-background}) directly in the global $n$-qubit Hilbert space, embedding each local stiffness contribution $\hat{K}_e$ into the full $2^n$-dimensional space to obtain $\mathcal{P}(K_e)$ at a cost of $O(n2^n)$, since the local sparsity of $\hat{K}_e$ is independent of $n$.

To estimate the total work at depth $k$, we combine the $O(n2^n)$ per-element decomposition cost with the scaling of $|T_{[k]}^\times|$ from Section~\ref{sec:asymptotic-scaling-of-cut-elts}:
\begin{equation}
    R_{\textrm{TPD}}(n, k) \asymp n2^n |T_{[k]}^\times|\asymp n 2^{\beta_d^- k + \beta_d^+ n},
\end{equation}
where $R_{\textrm{TPD}}(n, k)$ is the asymptotic complexity cost of processing all cut elements at
depth $k$ where the global label space is of size $2^n$. A detailed derivation of this per-depth cost, including the bound on nonzero Pauli terms, is provided in Appendix~\ref{app:full-space-tpd-runtime-derivation}

\subsection{Reduced-Space TPD Method}
\label{sec:reduced-space-tpd-method}

The second component of the PHASE assembly scheme addresses the exponential cost of 
computing $\mathcal{P}(K_e)$ in the full $2^n$-dimensional Hilbert space required by 
full-space TPD.

At depth $k$ of the recursive mesh hierarchy, all nodes contained in a subgraph share 
the same binary prefix of length $k$, referred to as the \emph{coarse address}
$\bar{q}\in \{0,1\}^k$. The remaining $p_k=n-k$ bits encode both the unresolved 
subgraph hierarchy below depth $k$ and the local element degrees of freedom, thereby
defining a \emph{local qubit subspace} of dimension $2^{p_k}$. The reduced space TPD 
method exploits this separation to perform all element decompositions in the local qubit subspace of size $p_k$ before lifting them up to global space of size $n$.

Since all nodes in a depth $k$ subgraph share the same coarse address, the element stiffness operator $\hat{K}_e$ acts nontrivially only on the remaining $p_k=n-k$ local qubits, and TPD can be applied entirely within this smaller space. For each cut element $e\in T_{[k]}^\times$, we can remove the leading $k$ bits of each node's global label to obtain a reduced-space operator
\begin{equation} \label{eq:k-tilde}
    \tilde{K}_e \in \R^{2^{p_k} \times 2^{p_k}}, \qquad p_k = n-k
\end{equation}
whose rows and columns correspond to the local bit labels within the subgraph. Applying
TPD (Section~\ref{sec:tpd-background}) to this smaller matrix gives
\begin{equation}
    \mathcal{P}(\tilde{K}_e) = \{(\lambda_e^\Lambda, \Lambda) \mid \Lambda \in \mathcal{L}_{p_k},\ \lambda_e^\Lambda \neq 0\}
\end{equation}
where $\mathcal{L}_{p_k} = \{I,X,Y,Z\}^{\otimes p_k}$ is the local $p_k$-qubit Pauli basis. This operation costs $O((n-k)2^{n-k})$ per element, since $\nu_e$ and local sparsity are independent of $n$. This reduced space method allows us to perform TPD at a smaller cost for high $k$ relative to full space TPD.

With the reduced-space decomposition $\mathcal{P}(\tilde{K}_e)$ in hand, it remains to lift each local Pauli representation back into the global $n$-qubit space by attaching a projection operator that enforces the coarse address $\bar{q}$ on the first $k$ bits. To re-embed the reduced Pauli representation $\mathcal{P}(\tilde{K}_e)$ into the global 
$n$-bit Hilbert space, we attach a tensor-product \emph{projection operator}
\begin{equation} \label{eq:proj-def}
    \Pi_{\bar{q}}^{(k)} = \frac{1}{2^k} \sum_{s \in \{0,1\}^k} \Phi_{\bar{q}}(s)\Xi_k (s),
    \qquad
    \Phi_{\bar{q}}(s) = (-1)^{\langle s, \bar{q} \rangle},
    \quad 
    \Xi_k (s) = \bigotimes_{i=1}^k Z_i^{s_i}
\end{equation}
where by convention $Z_i^0=I_i$, $Z_i^1=Z_i$, $I_i$ and $Z_i$ is an identity and a Pauli-$Z$ operation on the $i$-th qubit respectively, and the inner product 
$\langle s, \bar{q} \rangle = \sum_i s_i \bar{q}_i \; \operatorname{mod} 2$.

The projector $\Pi_{\bar{q}}^{(k)}$ enforces that the first $k$ qubits of any state coincide with the coarse address $\bar{q}$. In standard notation,
$$
|b_1 \dots b_n\rangle = |b_1 \dots b_k\rangle \otimes |b_{k+1} \dots b_n\rangle \longrightarrow \Pi_{\bar{q}}^{(k)} \otimes |b_{k+1} \dots b_n\rangle = |\bar{q}\rangle \otimes |b_{k+1} \dots b_n\rangle,
$$
so that each local contribution acts only within its designated subgraph. Each term in $\Pi_{\bar{q}}^{(k)}$ is a tensor product of $I$ and $Z$ operators weighted by the phase $(-1)^{\langle s, \bar{q}\rangle}$, and the sum over $s$ reproduces the rank-1 projector $|\bar{q}\rangle\langle\bar{q}|$ in Pauli form. Consequently, $\Upsilon_{\bar{q}}^{(k)}(\mathcal{P}(\tilde{K}_e))$ acts nontrivially only on qubits associated with the coarse address $\bar{q}$, while remaining block-diagonal across all other coarse addresses.

We define the lifting map $\Upsilon_{\bar{q}}^{(k)}$ acting on a reduced Pauli list by
\begin{equation}
    \Upsilon_{\bar{q}}^{(k)}\!\left(\mathcal{P}(\tilde{K}_e)\right) = \biguplus_{s \in \{0,1\}^k} \left\{\left(\tfrac{\Phi_{\bar{q}}(s)}{2^k}\,\lambda_e^\Lambda,\ \Xi_k(s) \otimes \Lambda\right) \mid (\lambda_e^\Lambda, \Lambda) \in \mathcal{P}(\tilde{K}_e)\right\}.
\end{equation}
The lifted global Pauli representation of element $e$ is therefore
\begin{equation}
\mathcal{P}(K_e) = \Upsilon_{\bar{q}}^{(k_e)}\!\left(\mathcal{P}(\tilde{K}_e)\right).
\end{equation}
A proof of correctness for $\Upsilon_{\bar{q}}^{(k)}$ is deferred to Appendix~\ref{app:projection-proof-of-correctness}.

Having established the lifting map, we now quantify the computational implications of performing TPD in the reduced space relative to the full $2^n$-dimensional embedding. Let $|T_{[k]}^\times|$ denote the number of elements at depth $k$, which scales as
$|T_{[k]}^\times | \asymp 2^{\beta_d n + \beta_d^- k}$. Decomposing each reduced 
space element and constructing its projector therefore incurs total cost
\begin{equation} \label{eq:scaling-of-local-tpd}
    R_{\textrm{TPD-Reduced}}(n,k) \asymp 
    (n - k) 2^{\beta_d^{+} n + k(\beta_d^{-} - 1)}
\end{equation}
A more detailed explanation for Equation~\ref{eq:scaling-of-local-tpd} can be found
in Appendix~\ref{app:reduced-space-tpd-asymptotic-complexity}.

\subsection{Global Aggregation via FWHT}
\label{sec:global-aggregation-via-fwht}

Once each reduced-space decomposition $\mathcal{P}(\tilde{K}_e)$ has been constructed, the next step is to apply the lifting map $\Upsilon_{\bar{q}}^{(k)}$ to embed these local operators into the global Hilbert space. A naive approach would apply each projector $\Pi_{\bar{q}}^{(k)}$ individually to every element and sum the resulting terms across subgraphs, incurring a per-element cost of $k2^k$ from inspecting $k$ bits for each of the $2^k$ bitmasks $\Xi_k(s)$, which becomes prohibitive at higher depths.

However, PHASE exploits the \emph{Fast Walsh-Hadamard Transform (FWHT)} (Section~\ref{sec:fwht-intro})
to perform this projection collectively at the level of an entire depth $k$.
The FWHT implements the action of all projectors
simultaneously by exploiting the orthogonality of their phase factors $\Phi_{\bar{q}}(s) = (-1)^{\langle s, \bar{q} \rangle}$,
which form a complete orthogonal basis over the binary address space 
$\{0, 1\}^{k}$. This allows all lifted contributions from the 
cut elements at depth $k$ to be aggregated in a single transform step for each local
Pauli pattern, rather than through elementwise accumulation.

To apply the FWHT, we first bin the reduced-space Pauli coefficients by coarse address and local pattern we define the \emph{binned coefficient vector}
\begin{equation}
    b^{(k, \Lambda)}(s) := 
    \sum_{\stackrel{e\in T_{[k]}^\times}{\bar{q}_e = s}}
    \lambda_e^\Lambda,
    \qquad
    s\in \{0, 1\}^k
\end{equation}
where $\lambda_e^\Lambda = 0$ for any $e$ such that $(\lambda_e^\Lambda, \Lambda) \notin \mathcal{P}(\tilde{K}_e)$ and whose $2^k$ components collect all coefficients associated with the
same coarse address $s$ at depth $k$. The vector $b^{(k, \Lambda)}$
thus represents the spatial distribution of a single Pauli 
pattern $\Lambda$ over the hierarchy at that level.

With the coefficients binned by coarse address, the FWHT replaces the explicit sum over $2^k$ addresses with a single structured transform, producing phase-weighted aggregates over all subgraphs at depth $k$ simultaneously. Applying the $k$-dimensional FWHT to $b^{(k, \Lambda)}$ yields
\begin{equation}
    (H^{(k)} b^{(k, \lambda)})(s) =
    \sum_{t\in \{0, 1\}^k} (-1)^{\langle s,t \rangle} 
    b^{(k, \Lambda)}(t) =
    \sum_{e \in T_{[k]}^\times} (-1)^{\langle s, \bar{q}_e \rangle} 
    \lambda_e^\Lambda.
\end{equation}
Define the \emph{normalized spectral coefficient}
\begin{equation}
    C^{(k, \Lambda)}(s) = \frac{1}{2^k}
    \sum_{e \in T_{[k]}^\times} \Phi_{\bar{q}_e}(s) \Lambda_e^\Lambda,
    \qquad \Phi_{\bar{q}}(s) = (-1)^{\langle s, \bar{q} \rangle}
\end{equation}
Each coefficient $C^{(k, \Lambda)}(s)$ represents the global, 
phase-weighted sum of all local contributions carrying the Pauli 
pattern $\Lambda$ at depth $k$. In this sense, the FWHT
serves as a discrete Fourier transform over the binary address space,
converting spatially localized data into its hierarchical spectral
representation.

These spectral coefficients assemble directly into the depth-$k$ Pauli aggregate by pairing each $C(k, \Lambda)(s)$ with its corresponding global Pauli term $\Xi_k(s) \otimes \Lambda$. Ranging over all local patterns $\Lambda \in \mathcal{L}_{p_k}$ and all coarse addresses $s \in \{0,1\}^k$, these coefficients populate the merged tuple list for depth $k$:
\begin{equation}
\biguplus_{e\in T_{[k]}^\times} \mathcal{P}(K_e) = \left\{\left(C^{(k,\Lambda)}(s),\ \Xi_k(s) \otimes \Lambda\right) \ \middle|\ \Lambda \in \mathcal{L}_{p_k},\ s \in \{0,1\}^k,\ C^{(k,\Lambda)}(s) \neq 0\right\},
\end{equation}
with $\Xi_k(s)$ as defined in Eq~\eqref{eq:proj-def}. This identity expresses the depth-$k$ aggregate as a structured merge
of Kronecker products between coarse-scale $Z$-bitmask operators and
fine-scale Pauli patterns, with zero-coefficient terms excluded by the tuple-set convention.

We now account for the total cost of using FWHT aggregation. Evaluating all coefficients $C^{(k, \Lambda)}(s)$ over $\Lambda \in \mathcal{L}_k$ requires performing 
$4^{p_k} = 4^{n-k}$ FWHTs of length $2^k$, giving a total cost
\begin{equation}\label{eq:r-fwht}
    R_{\mathrm{FWHT}}(n,k) \asymp k 2^k 4^{n-k} \asymp k 2^{2n-k}.
\end{equation}
This aggregation term, combined with the reduced-space TPD cost from Eq.~\eqref{eq:scaling-of-local-tpd} forms the complete per-level asymptotic runtime complexity of the FWHT-based method:
\begin{equation}
    R_{\textrm{TPD-R}}(n,k) + R_{\textrm{FWHT}}(n,k) \asymp
    (n-k) 2^{\beta_d^+ n + k(\beta_d^- - 1)}
    + k2^{2n-k}.
\end{equation} 
The first term arises from the global aggregation through the FWHT,
while the second reflects the cost of the local Pauli decomposition.
This formulation provides the foundation for the hybrid asymptotic 
model developed in Section~\ref{sec:asymptotic-runtime-model}.


\section{Asymptotic Complexity Analysis}
\label{sec:asymptotic-runtime-model}

At depth $k$ of the recursive partition tree, PHASE may assemble cut elements using either (1)~{Full-space Tensorized Pauli Decomposition (TPD)}, operating in the global $n$-qubit Hilbert space, or (2)~{Reduced-space TPD followed by FWHT aggregation}, operating in a local subspace of dimension $n-k$ and lifting contributions globally via Walsh-Hadamard mixing. These two regimes exhibit complementary asymptotic behavior. Full-space TPD incurs a fixed exponential cost in $n$ per element, but benefits from the fact that the number of cut elements grows slowly for small $k$. In contrast, the FWHT-based method becomes increasingly efficient as $k$ grows, since the local Pauli decompositions shrink exponentially in size, while global aggregation is amortized across subdomains.

Motivated by this trade-off, we consider a \emph{hybrid depth threshold $j$}, defined such that depths $k<j$ are processed using full-space TPD, and depths $k \ge j$ are processed using the FWHT-based reduced-space method. This partition allows us to combine the two asymptotic regimes into a single model and determine the optimal transition depth $j^\star$.

\subsection{Simplified Global Asymptotic Runtime Complexity Model}
As elaborated in Sections~\ref{sec:full-space-tpd}-\ref{sec:global-aggregation-via-fwht}, the dominant per-depth costs are: 

\textbf{Full-space TPD} (for $k < j$)
\begin{equation} \operatorname{R_{TPD}}(n,k) = O(n 2^{\beta^+_d n + \beta_d^- k}), \end{equation}
where $\beta_d = {d-1}/{d}$, $\beta_d^- = {1}/{d}$, and $\beta_d^+={2d-1}/{d}$. 

\textbf{FWHT aggregation} (for $k \geq j$)
\begin{equation} \operatorname{R_{FWHT}}(n,k) = O(k 2^{2n-k}). \end{equation}

The reduced-space TPD cost is asymptotically dominated by the FWHT aggregation term for all fixed $d$, and is therefore omitted from the leading-order model (see Appendix~\ref{app:hybrid-asymptotic-complexity-model}). Summing over depths yields the total complexity
\begin{equation}
R(n, j) = \sum_{k=0}^{j-1} n 2^{\beta_d^+ n + \beta_d^- k} + \sum_{k=j}^{n}k 2^{2n-k}.
\end{equation}
Both sums are geometrically dominated by their largest terms. Evaluating these leading contributions gives a simplified expression
\begin{equation}\label{eq:full-asymptotic-runtime}
R(n,j) = n 2^{\beta_d^+ n + \beta_d^- j} + j 2^{2n-j}.
\end{equation}
The first term increases monotonically with $j$, reflecting the rising cost of extending full-space TPD into the hierarchy. The second term decreases monotonically with $j$, reflecting the shrinking cost of FWHT aggregation as local subspaces collapse.

\subsection{Optimal Transition Depth and Dimensional Scaling}
\label{sec:optimal-transition-depth}

The optimal transition depth $j^\star$ minimizes $R(n,j)$. Since both terms in (\ref{eq:full-asymptotic-runtime}) are exponential in $n$, the minimum occurs when their exponents are asymptotically balanced:
\begin{equation}
n 2^{\beta_d^+ n + \beta_d^- j} \asymp j 2^{2n-j}.
\end{equation}
Neglecting logarithmic prefactors and equating exponents yields
\begin{equation}
\beta_d^+ n + \beta_d^- \asymp 2n-j,
\end{equation}
which gives
\begin{equation}
j^\star \asymp \frac{\beta_d^-}{1+\beta_d^-} n = \frac{n}{d+1}
\end{equation}
Substituting $j^\star$ into (\ref{eq:full-asymptotic-runtime}), both terms scale with $2^{\gamma_d n}$, and the total complexity at the optimal transition depth is therefore
\begin{equation}
W(n,j^\star) \asymp n 2^{\gamma_d n}, \qquad \gamma_d = 2-\frac{1}{d+1}.
\end{equation}
This result demonstrates that PHASE achieves a {dimension-dependent reduction in the exponential scaling exponent} relative to the standard Tensor Pauli decomposition, with the strongest gains in low spatial dimensions.

\subsection{Unbalanced Partition Analysis}
\label{sec:unbalanced-partition-analysis}

The analysis above assumes approximately balanced recursive partitions. In practice, geometric or graph-based separators may produce systematically or stochastically unbalanced splits, which affect both the depth of the hierarchy and the relative cost of full-space and reduced space assembly. We therefore extend the asymptotic runtime complexity model to account for bounded imbalance.

Consider a subgraph of size $M$ split into children of sizes $(M_L, M_S)$, with
\begin{equation}
    M_L \geq M_S, \qquad M_L + M_S = M,
\end{equation}
and define the \emph{imbalance ratio}
\begin{equation}
    \delta := \frac{M_L}{M} \in [1/2,1).
\end{equation}
It is convenient to parameterize imbalance by the associated \emph{logarithmic shrink factor}
\begin{equation}
    \eta(\delta) := -\log_2 \delta \in (0,1]
\end{equation}
Balanced bisection corresponds to $\eta=1$, while increasingly skewed partitions correspond to $\eta \downarrow 0$.

Incorporating this imbalance into the hybrid TPD-FWHT model yields three asymptotic regimes, determined by the relative decay rates of reduced-space support. The resulting total complexity satisfies
\begin{equation} \label{eq:gamma-def}
W(n) \asymp 
\begin{cases}
n2^{\gamma(\eta,d)n}, \quad \eta > 1/2 \\
n^{2}2^{2n}, \quad \eta = 1/2, \qquad\qquad \gamma(\eta, d) = 2- \frac{2\eta - 1}{1+d(2\eta - 1)} \\
n 2^{n/\eta}, \quad 0 < \eta < 1/2
\end{cases}
\end{equation}
The regime $\eta > 1/2$ corresponds to a \emph{mild imbalance}, for which the FWHT contribution decays with depth and an optimal switching depth exists between full-space TPD and reduced-space aggregation. In this regime, the exponent $\gamma(\eta,d)$ interpolates smoothly between the balanced result $\gamma(1,d)=2-\frac{1}{d+1}$ and the naive $2^{2n}$ scaling as $\eta \downarrow 1/2$. The boundary case $\eta=1/2$ (equivalently, $\delta \approx 0.71)$ represents a \emph{knife-edge regime} in which FWHT costs no longer decay with depth, eliminating the benefit of deep hierarchical aggregation. For $\eta>1/2$, corresponding to \emph{strongly unbalanced partitions}, FWHT costs grow with depth and dominate the runtime. In this case, the optimal strategy is to avoid hierarchical aggregation altogether, and complexity deteriorates to $O(n 2^{n/\eta})$.

These results show that PHASE is robust to moderate imbalance but degrades predictably when recursive partitions become too skewed. Full derivations, including depth bounds, cut element estimates under imbalance, and asymptotic balancing arguments are provided in Appendix~\ref{app:unbalanced-partition-analysis}


\section{Numerical Verification}
\label{sec:numerical-validation}

We present numerical experiments validating the asymptotic complexity bounds derived in Section~\ref{sec:asymptotic-runtime-model}. All experiments measure wall-clock CPU time for the PHASE assembly procedure on finite element meshes of varying size. Recursive mesh partitioning was performed using the geometric separator algorithm of~\cite{miller1998geometric}. For each experiment, the partition imbalance $\delta$ is recorded at every cut, and the worst-case and average-case logarithmic shrink factors $\nu_\star = - \log_2 \delta_\star$ and $\nu_{\text{avg}} = -\log_2 \delta_{\text{avg}}$ are used to compute the corresponding theoretical scaling exponents $\gamma_\star$ and $\gamma_{\text{avg}}$ via Eq.~\eqref{eq:gamma-def}. Empirical scaling exponents $\hat{\gamma}$ are extracted by fitting a line to $\log_2(T/n)$ versus $n$ so that the slope is $\gamma$ (see Eq.~\eqref{eq:gamma-def}), and are compared against theoretical predictions. 

For each mesh instance, PHASE is executed at every admissible transition depth $j=0,\dots,n$ with three timing trials per depth; the median runtime at each depth is recorded and the optimal transition depth $j_{\text{obs}}^\star$ is identified as the minimizer. Three additional timing trials are then conducted at $j_{\text{obs}}^\star$ and the mean and standard deviation of these trials constitute the reported data point. Since $n$ depends on the number of degrees of freedom produced by the mesh generator and the depth of the recursive partition, which is random, the values of $n$ represented in each experiment are not prescribed but instead arise from the mesh geometry and bisector realization. Consequently, certain values $n$ may be absent from a given experiment. Where multiple mesh instances produce the same effective qubit register length $n$, their runtimes are pooled and the sample mean and standard deviation are reported, which accounts for variation in error bar size visible across data points.

\begin{figure}[htbp]
  \centering
  
  \makebox[0.03\textwidth][l]{\bfseries A}%
  \includegraphics[width=0.14\textwidth]{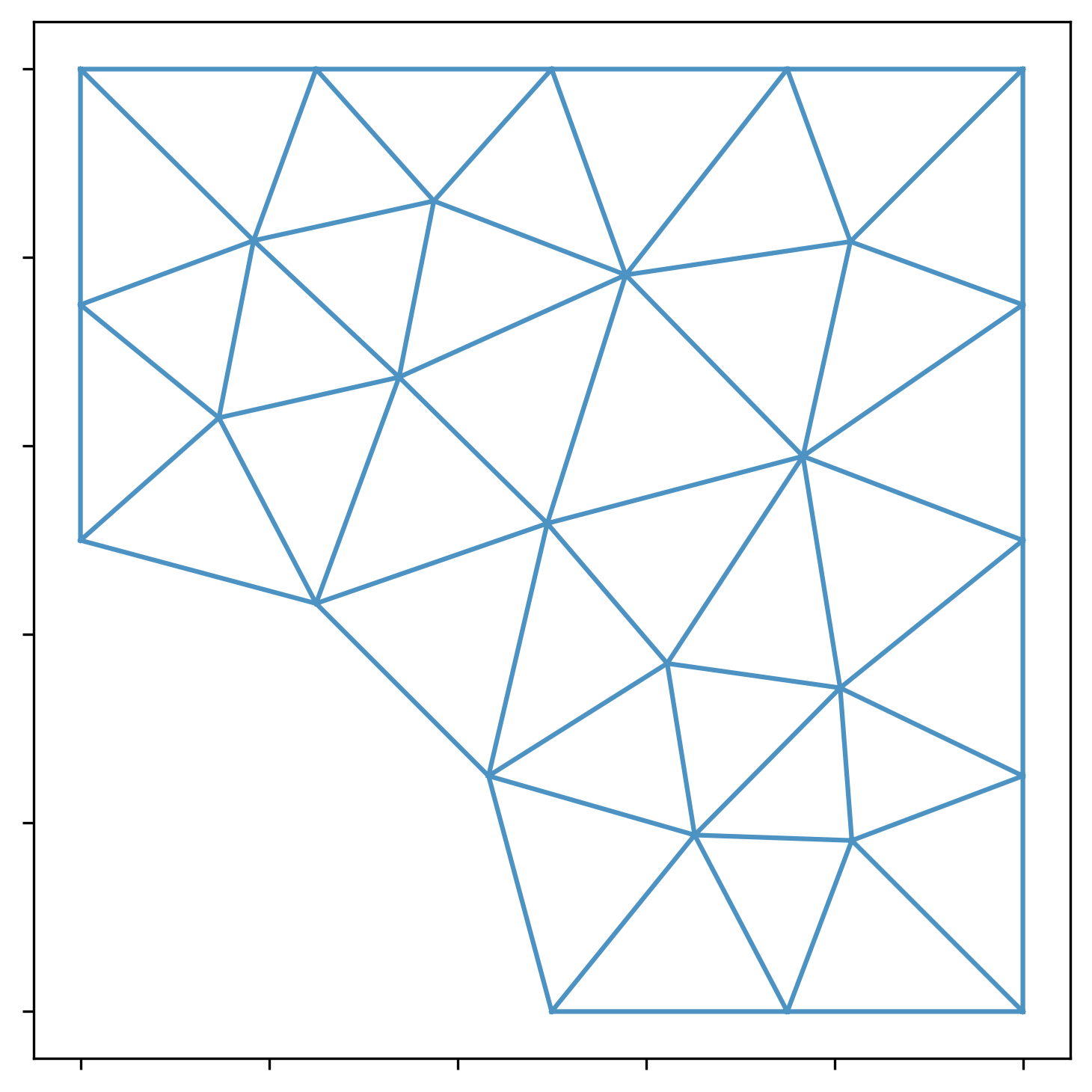}\hfill
  \includegraphics[width=0.14\textwidth]{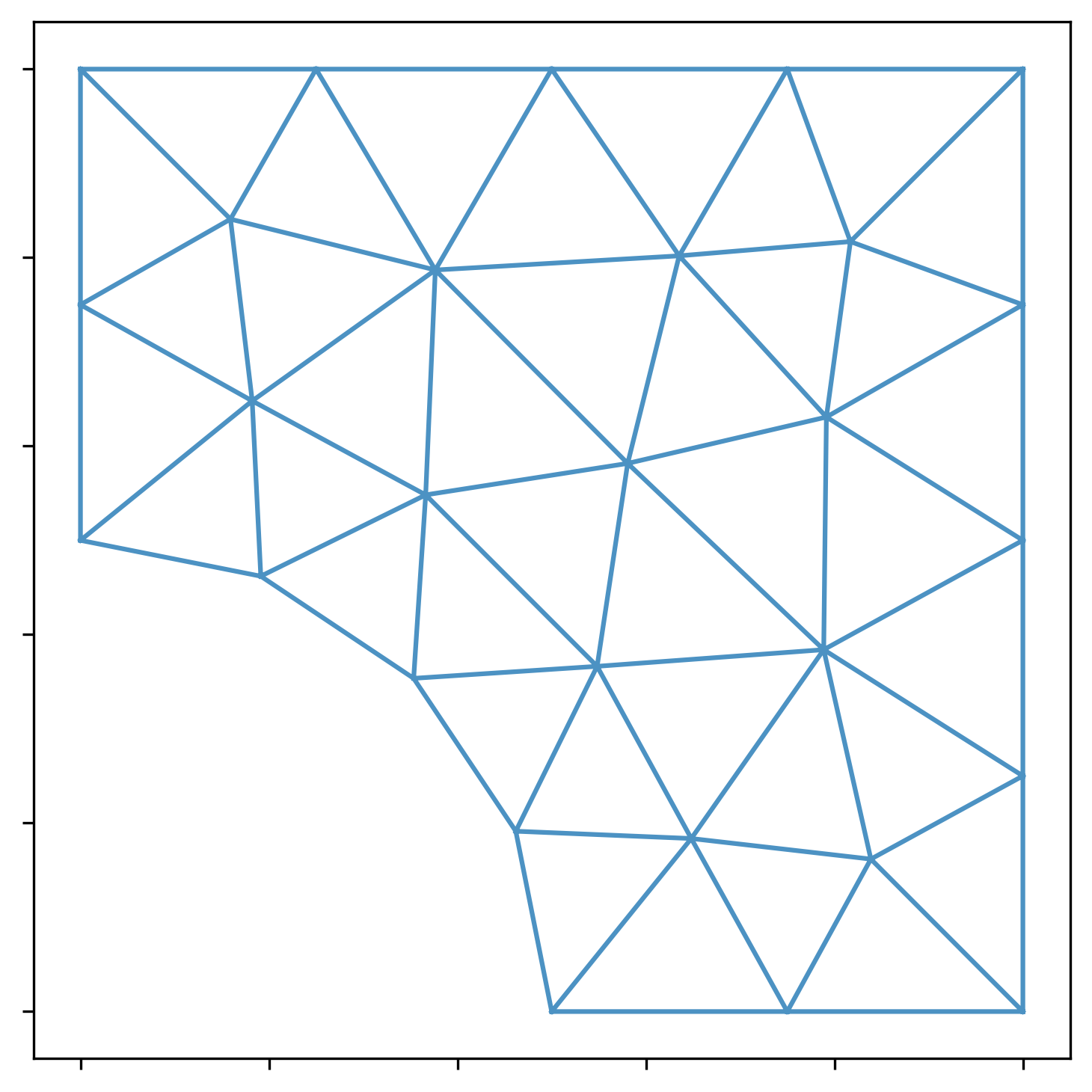}\hfill
  \includegraphics[width=0.14\textwidth]{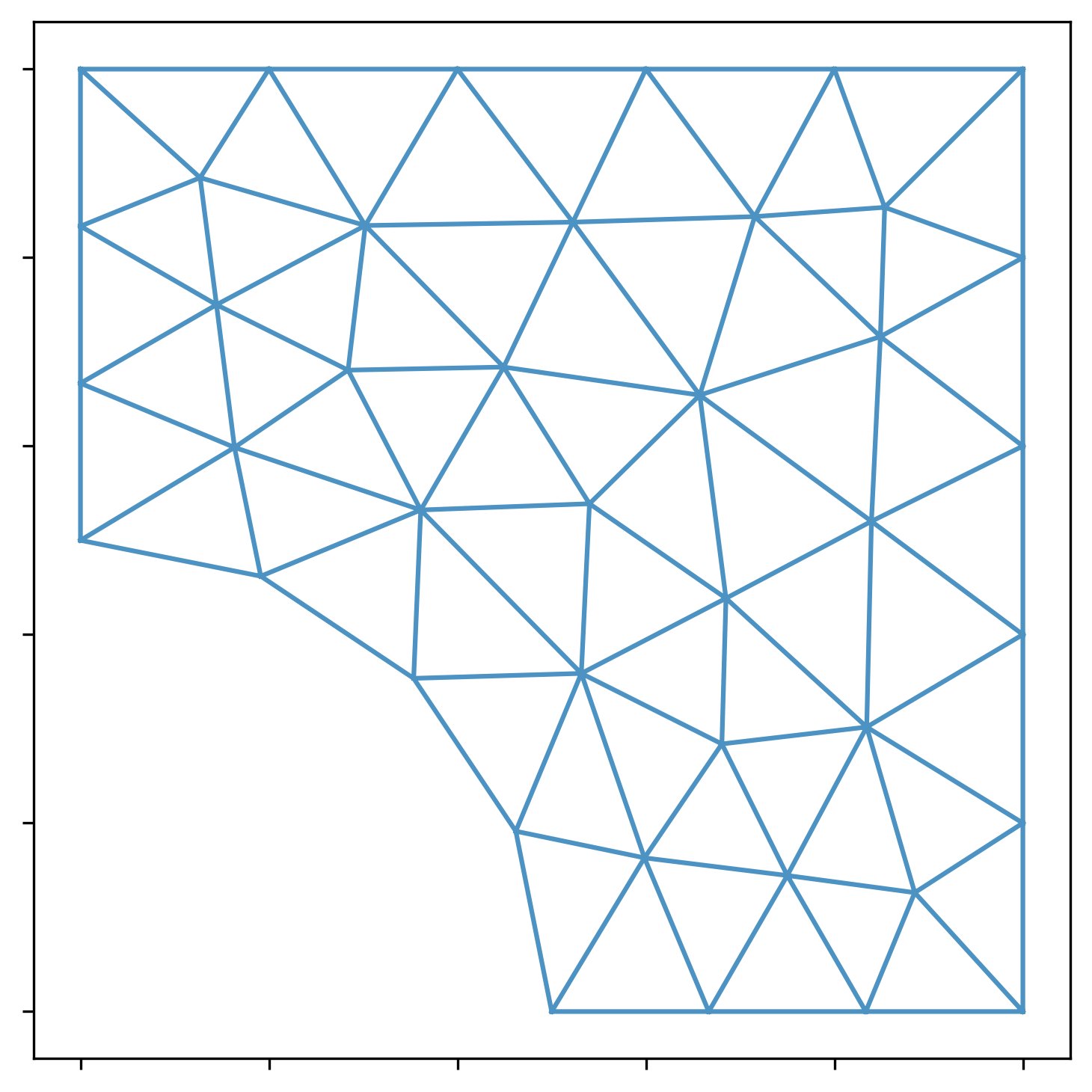}\hfill
  \includegraphics[width=0.14\textwidth]{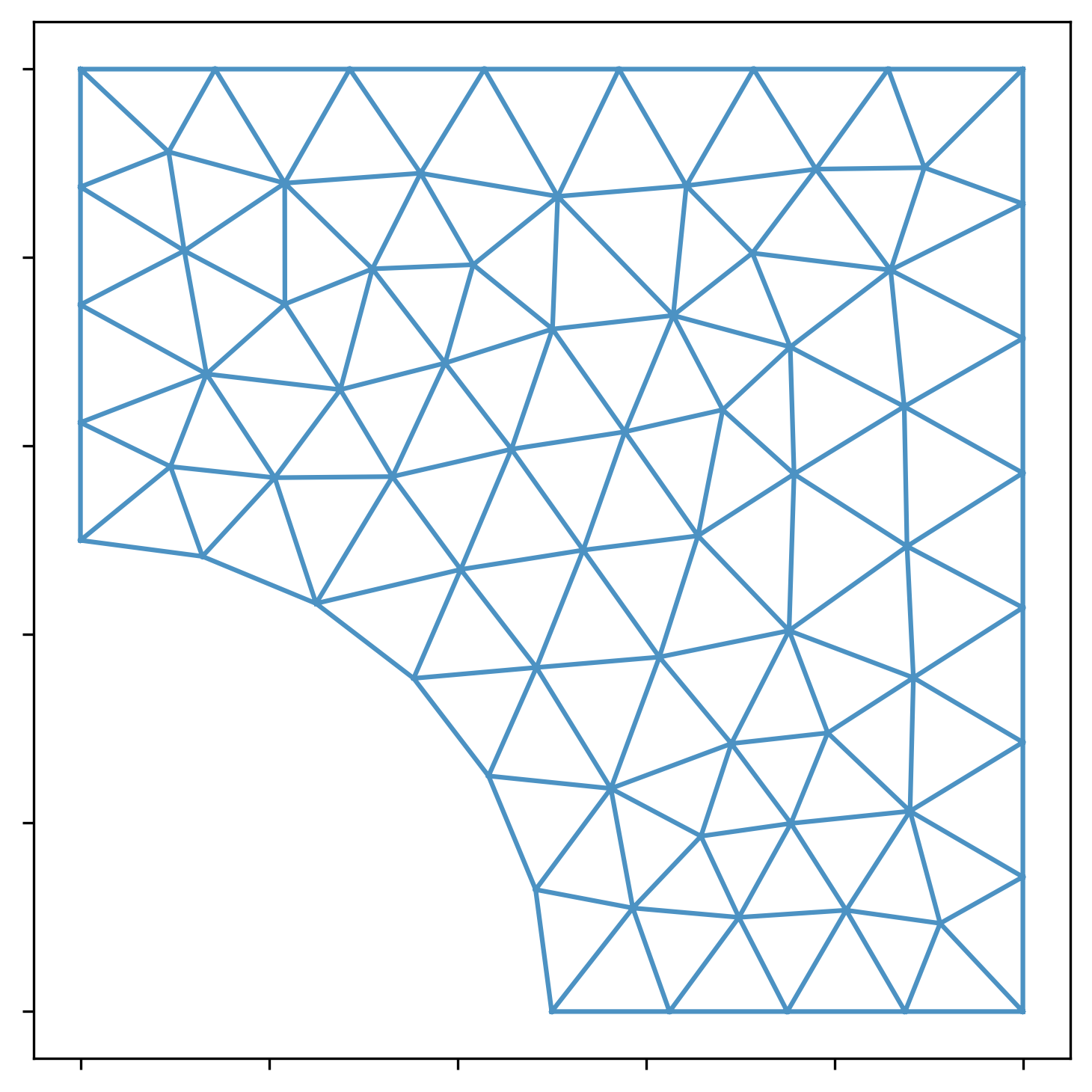}\hfill
  \includegraphics[width=0.14\textwidth]{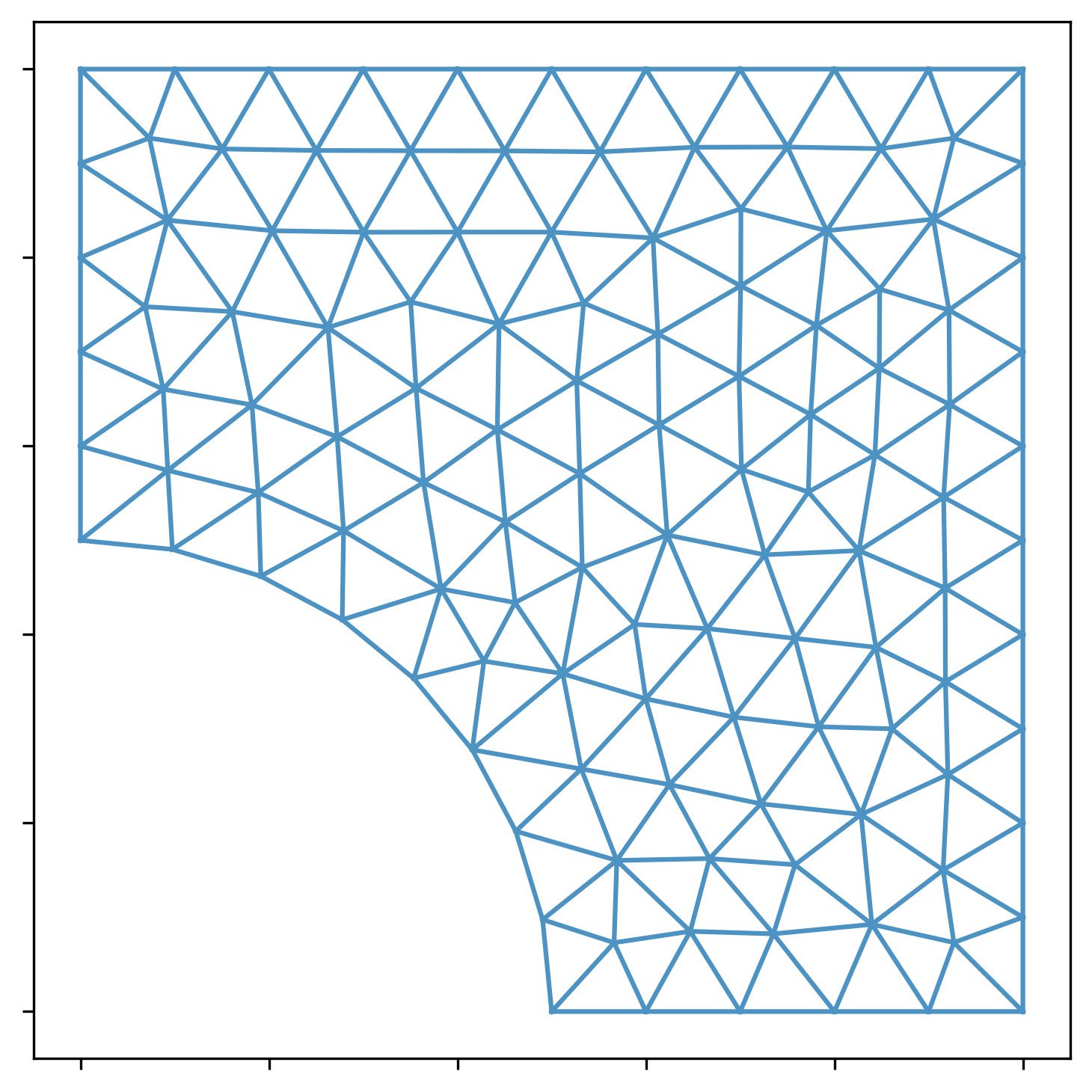}\hfill
  \includegraphics[width=0.14\textwidth]{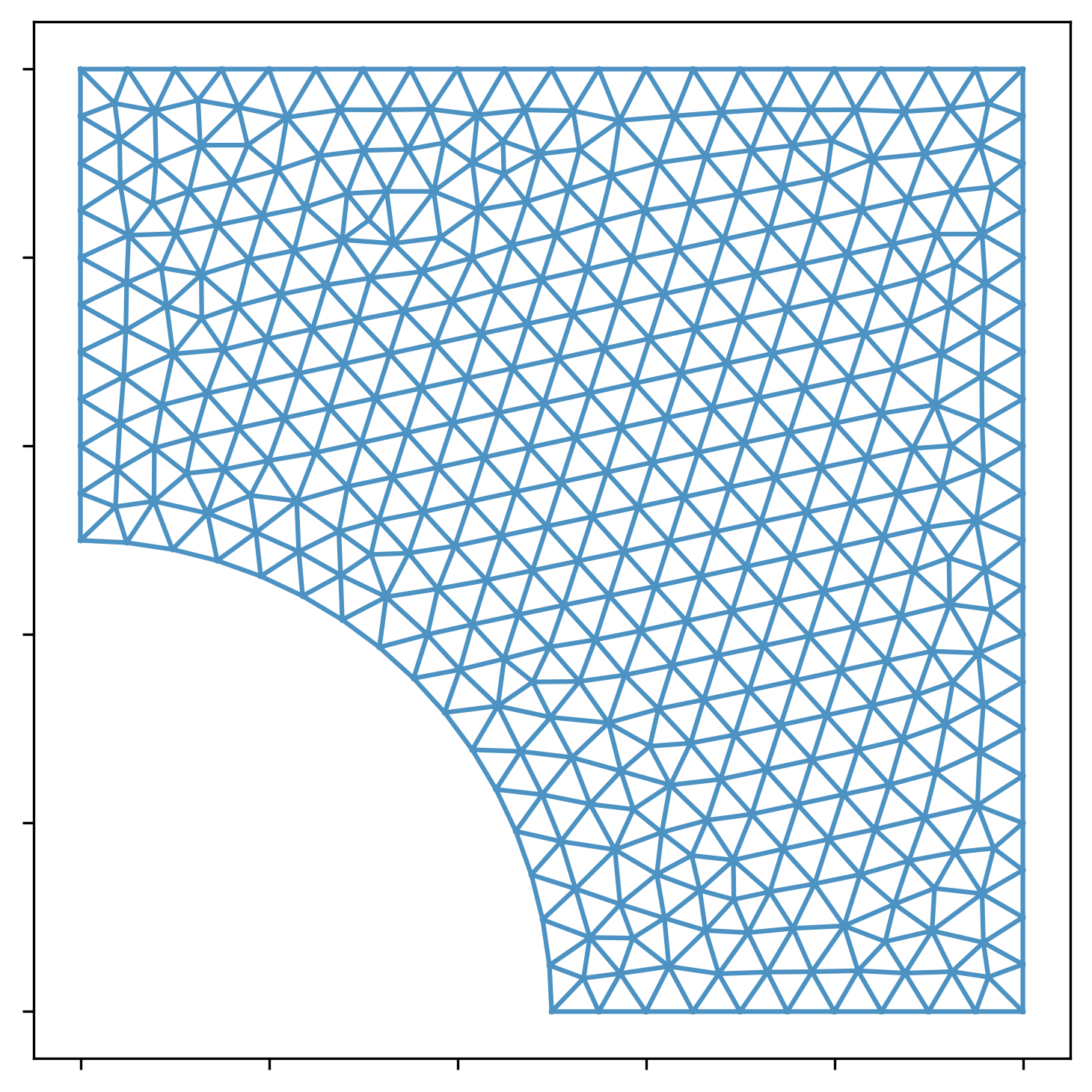}
  
  \vspace{0.5em}
  
  \makebox[0.03\textwidth][l]{\bfseries B}%
  \includegraphics[width=0.14\textwidth]{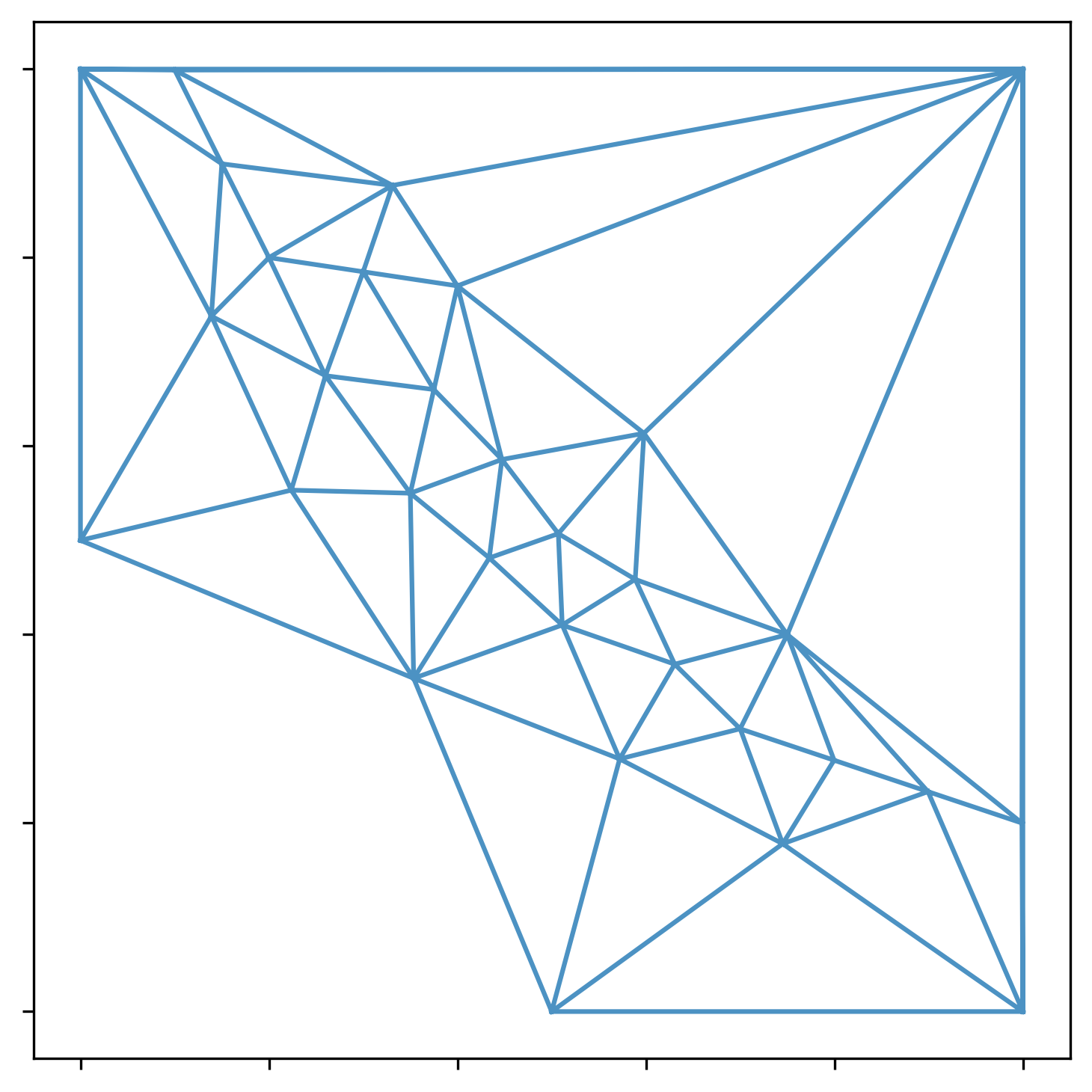}\hfill
  \includegraphics[width=0.14\textwidth]{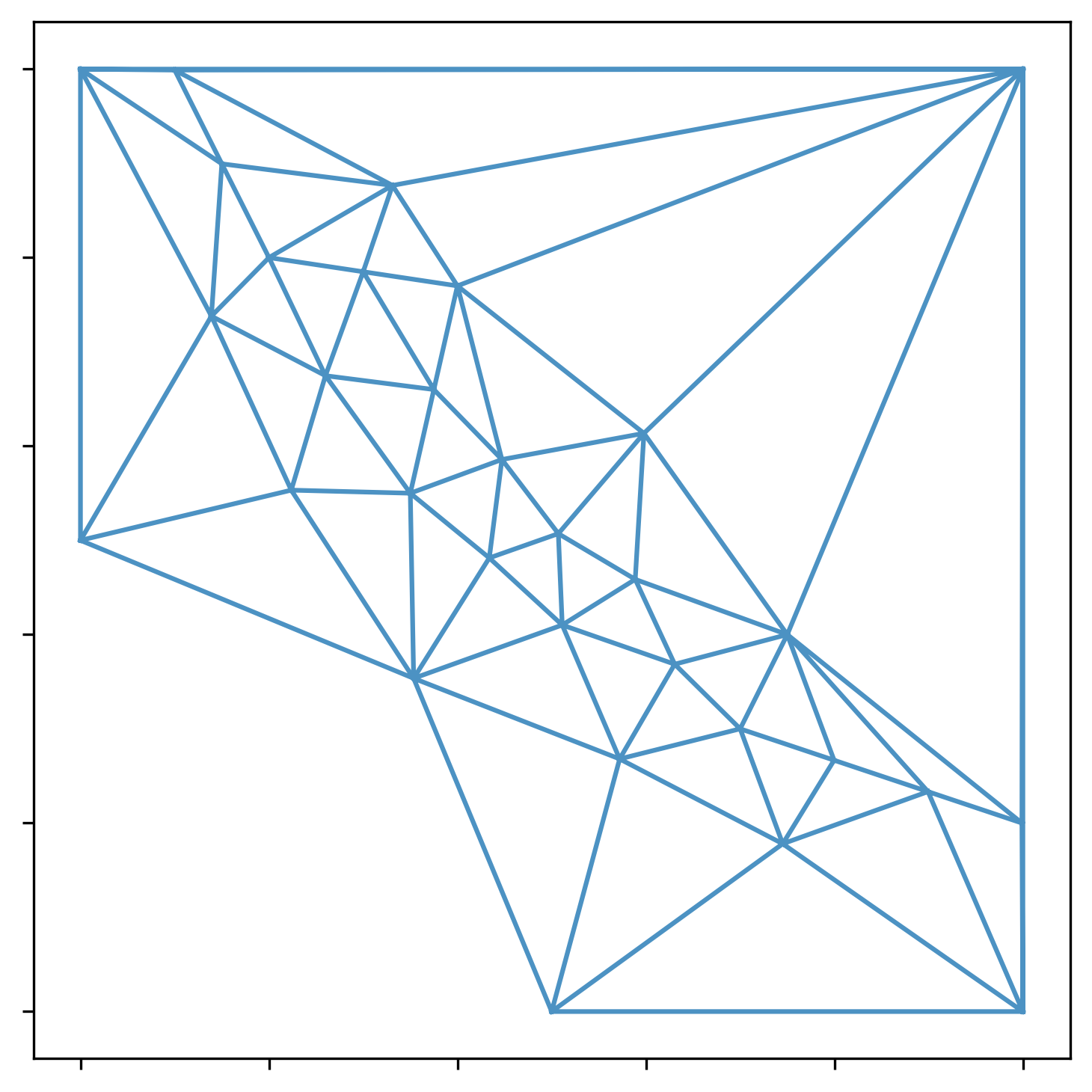}\hfill
  \includegraphics[width=0.14\textwidth]{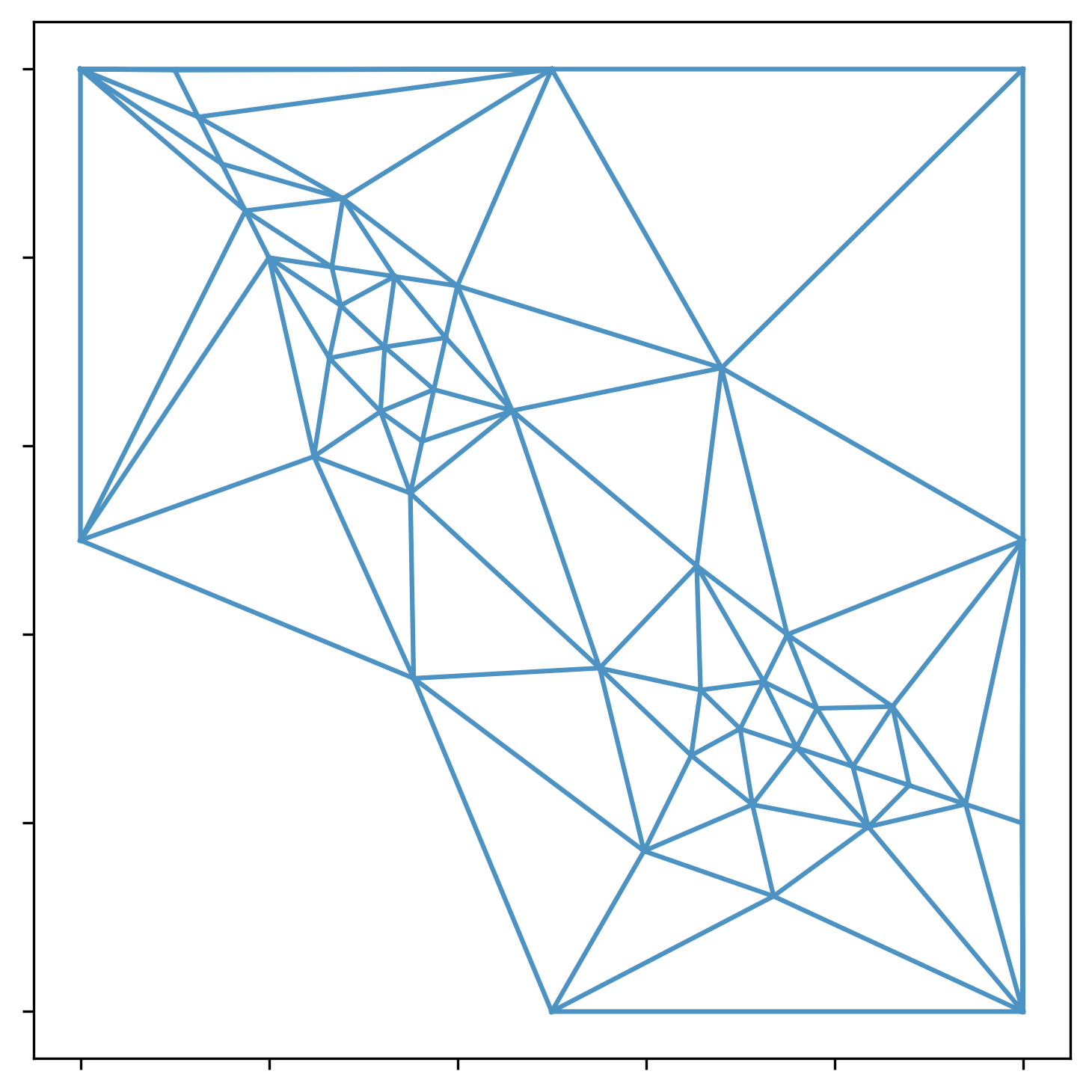}\hfill
  \includegraphics[width=0.14\textwidth]{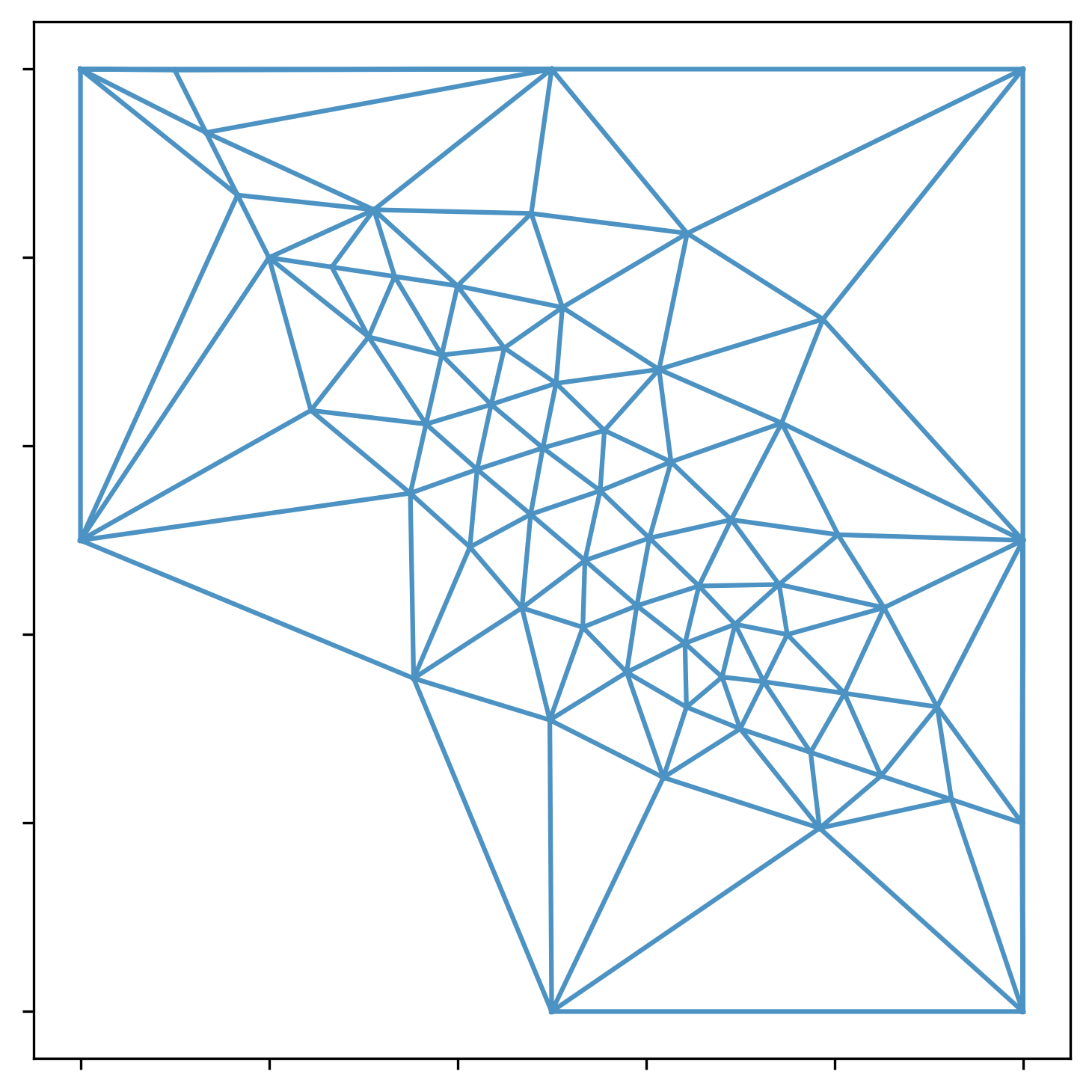}\hfill
  \includegraphics[width=0.14\textwidth]{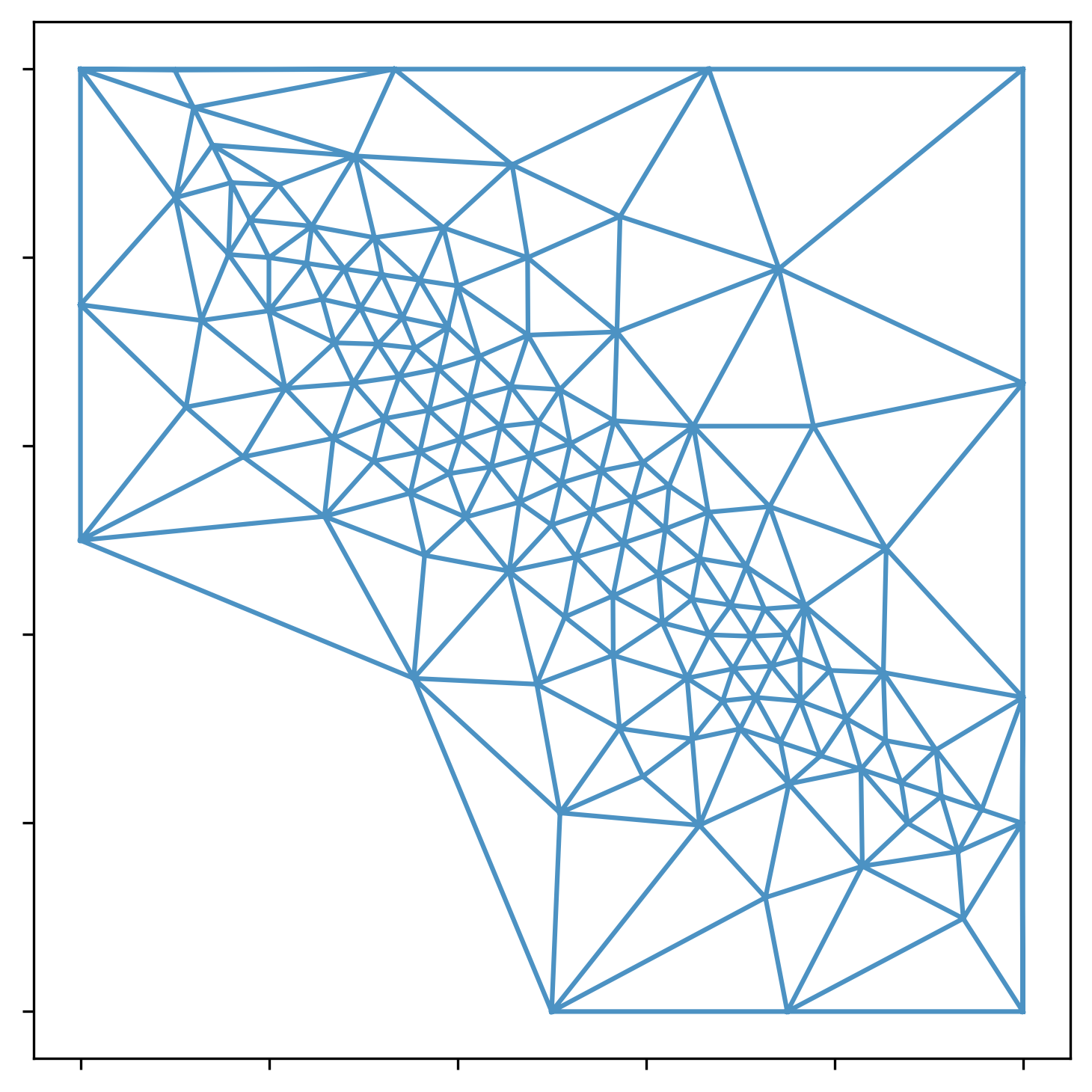}\hfill
  \includegraphics[width=0.14\textwidth]{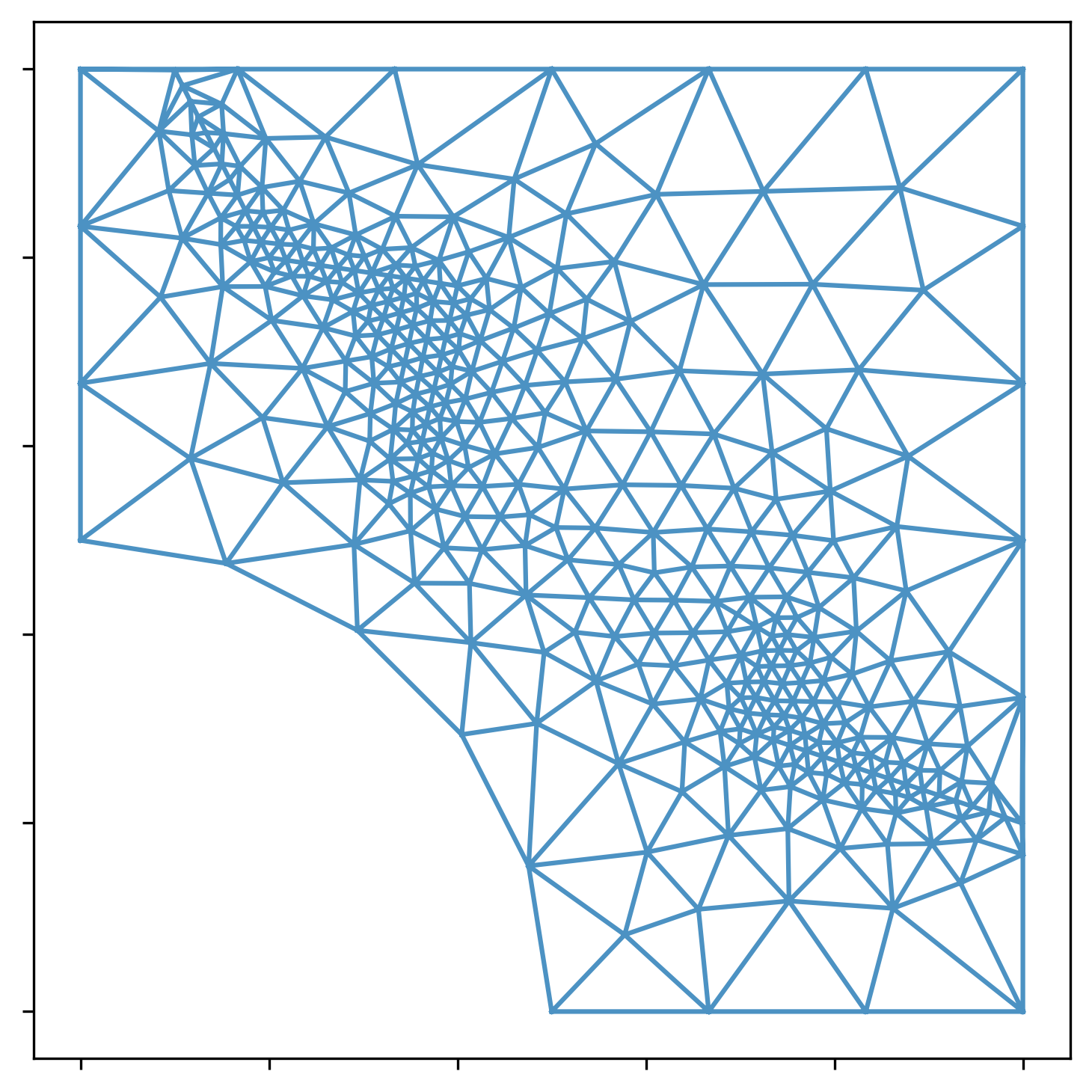}
  
  \vspace{0.5em}
  
  \makebox[0.03\textwidth][l]{\bfseries C}%
  \includegraphics[width=0.14\textwidth]{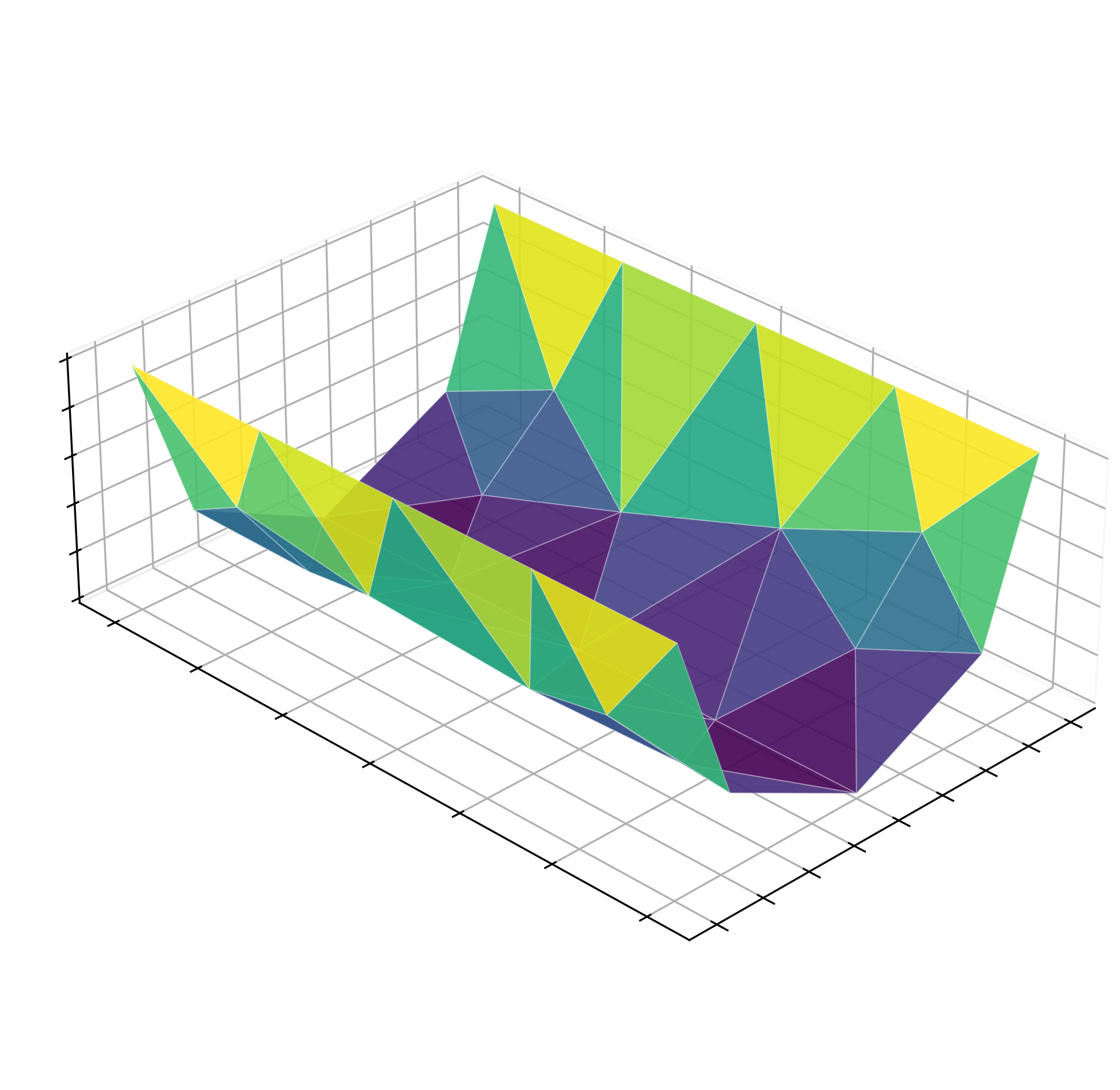}\hfill
  \includegraphics[width=0.14\textwidth]{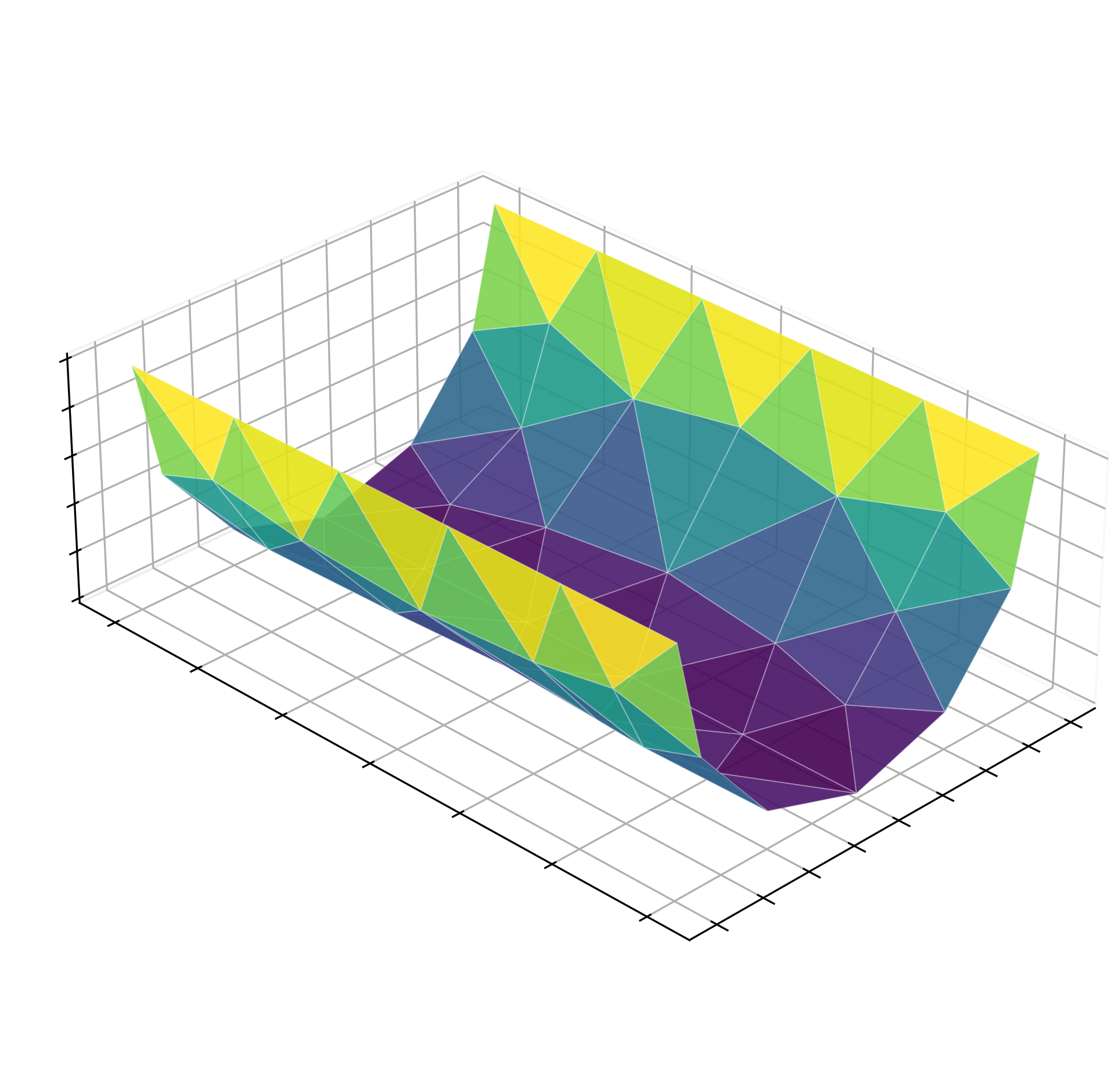}\hfill
  \includegraphics[width=0.14\textwidth]{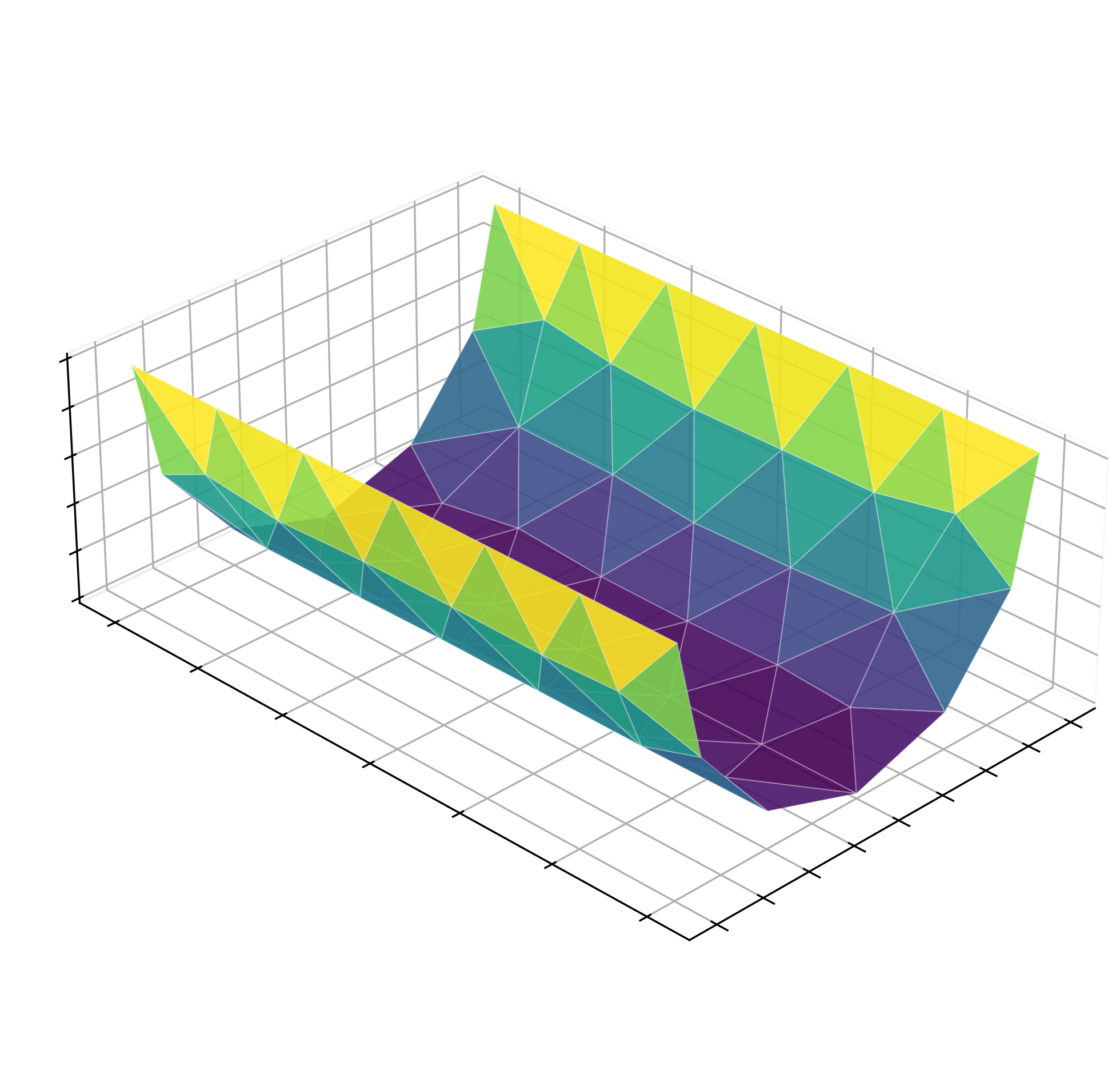}\hfill
  \includegraphics[width=0.14\textwidth]{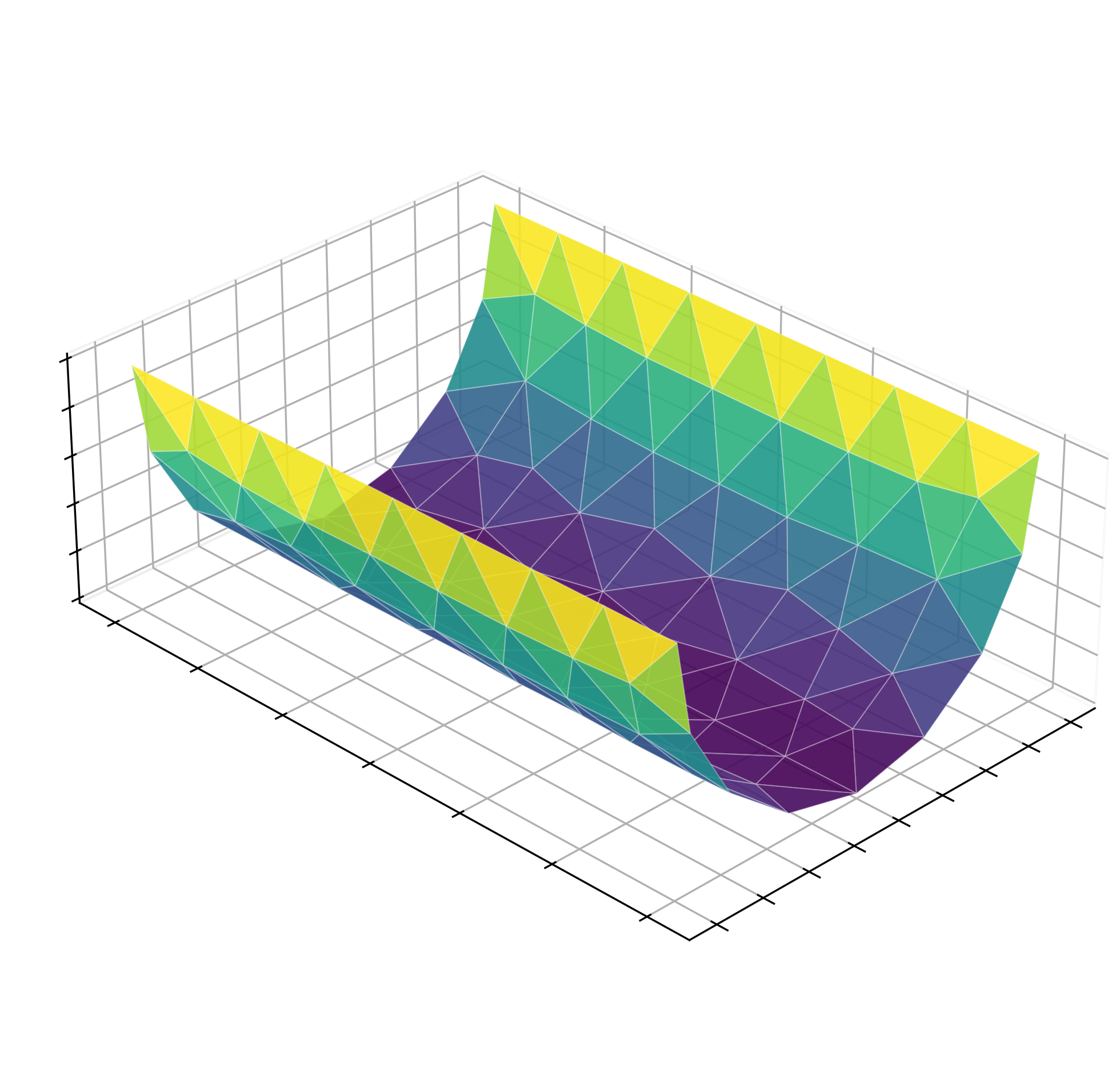}\hfill
  \includegraphics[width=0.14\textwidth]{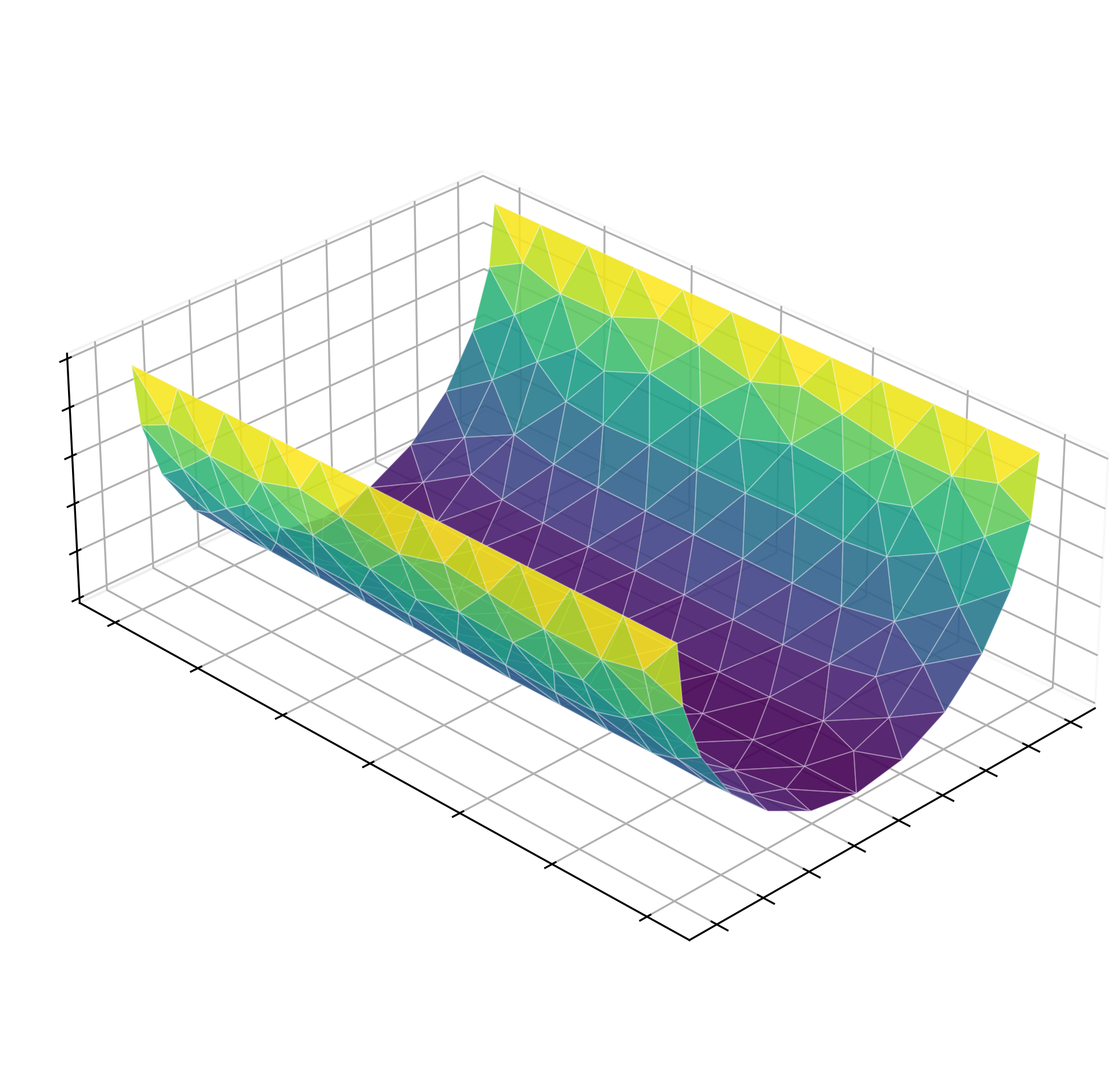}\hfill
  \includegraphics[width=0.14\textwidth]{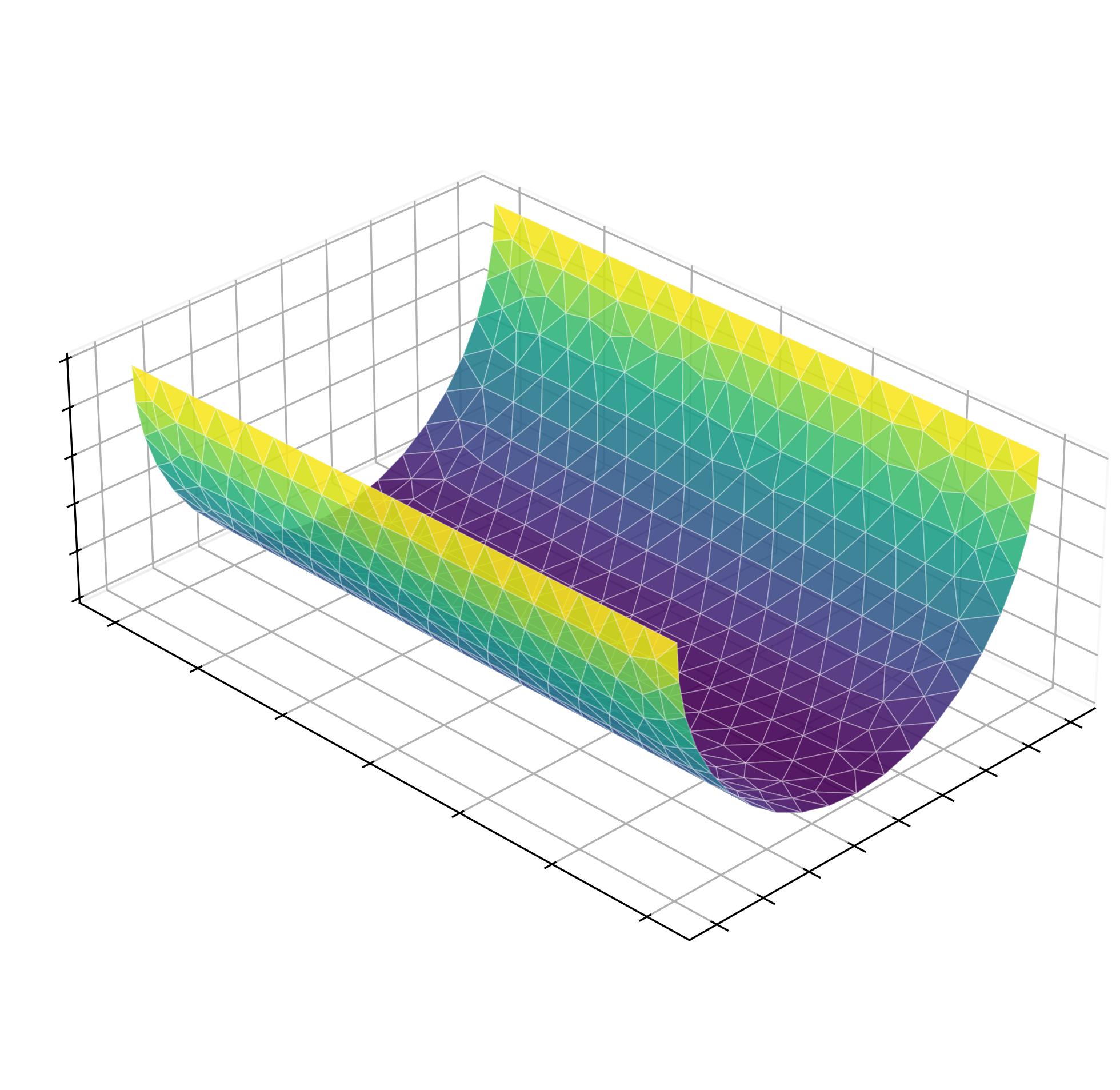}
  
  \caption{Meshes A, B and C at varying levels of refinement for empirical PHASE runtime testing.}
  \label{fig:all_mesh_plots}
\end{figure}

We consider three finite element meshes of varying geometric complexity divided into 6 different refinement levels as illustrated in Figure~\ref{fig:all_mesh_plots}. Mesh A is an unstructured but quasi-uniform triangular mesh of a square plate with a circular interior cutout, representing a smooth but non-convex domain; this mesh produces consistently moderate partition imbalance and serves as a primary validation case for the Region I scaling bound. Mesh B is identical to Mesh A except for two cracks emanating inward from the north and east faces of the domain. The second mesh stands as a standard benchmark geometry in computational fracture mechanics; the geometric irregularity introduced by the crack produces occasional moderate to highly imbalanced cuts, making this a natural test case for the imbalance sensitivity analysis of Section~\ref{sec:unbalanced-partition-analysis}. Mesh C is a halfpipe shell. This is a two-dimensional surface mesh embedded in $\mathbb{R}^3$, included to demonstrate that the scaling exponent $\gamma_d$ is governed by the topological dimension $d=2$ of the mesh rather than the dimension of ambient space.

\begin{figure}[t]
    \centering
    \begin{subfigure}[b]{0.48\textwidth}
        \centering
        \includegraphics[width=\textwidth]{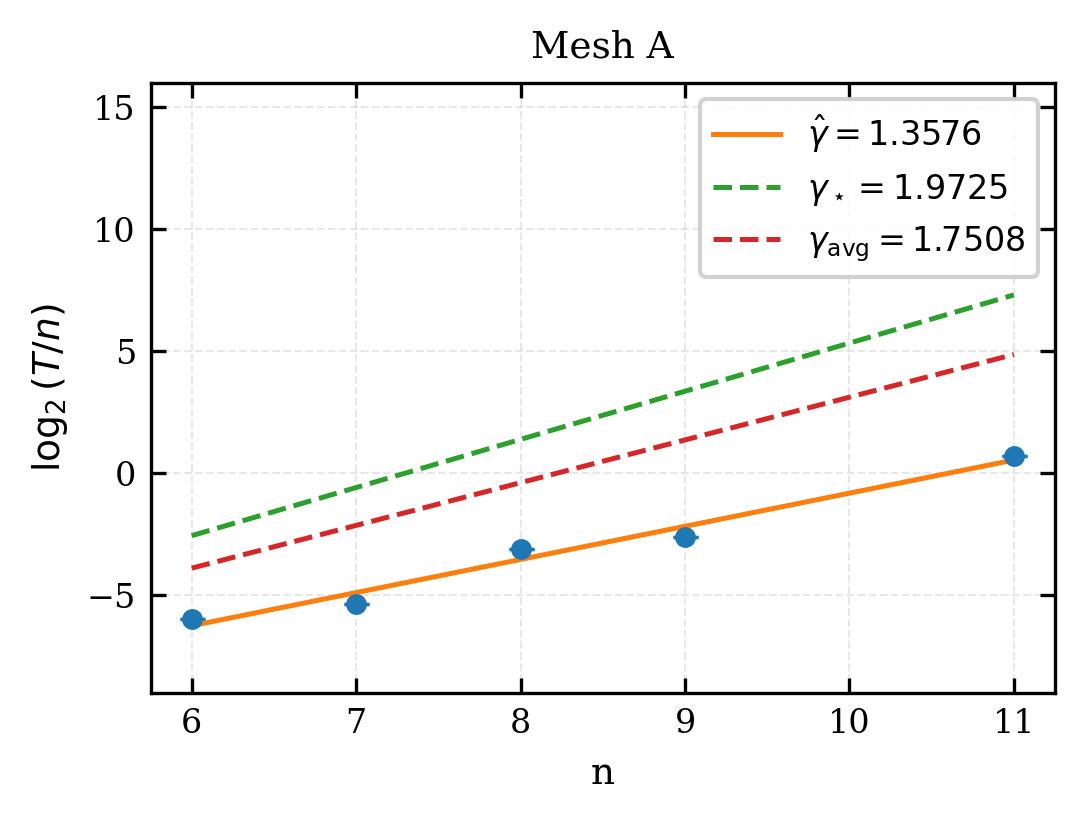}
        \caption{Mesh A: circular cutout plate}
        \label{fig:plot1}
    \end{subfigure}
    \hfill
    \begin{subfigure}[b]{0.48\textwidth}
        \centering
        \includegraphics[width=\textwidth]{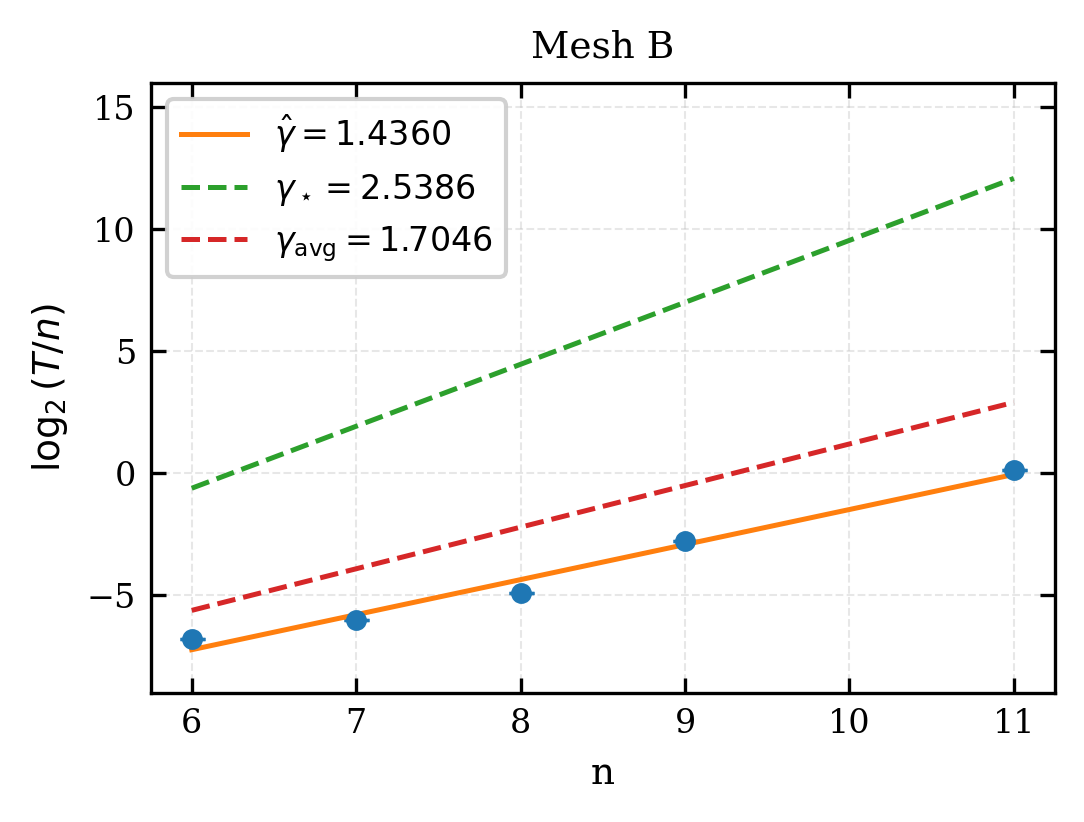}
        \caption{Mesh B: edge-cracked plate}
        \label{fig:plot2}
    \end{subfigure}
    
    \vspace{0.5cm}
    
    \begin{subfigure}[b]{0.48\textwidth}
        \centering
        \includegraphics[width=\textwidth]{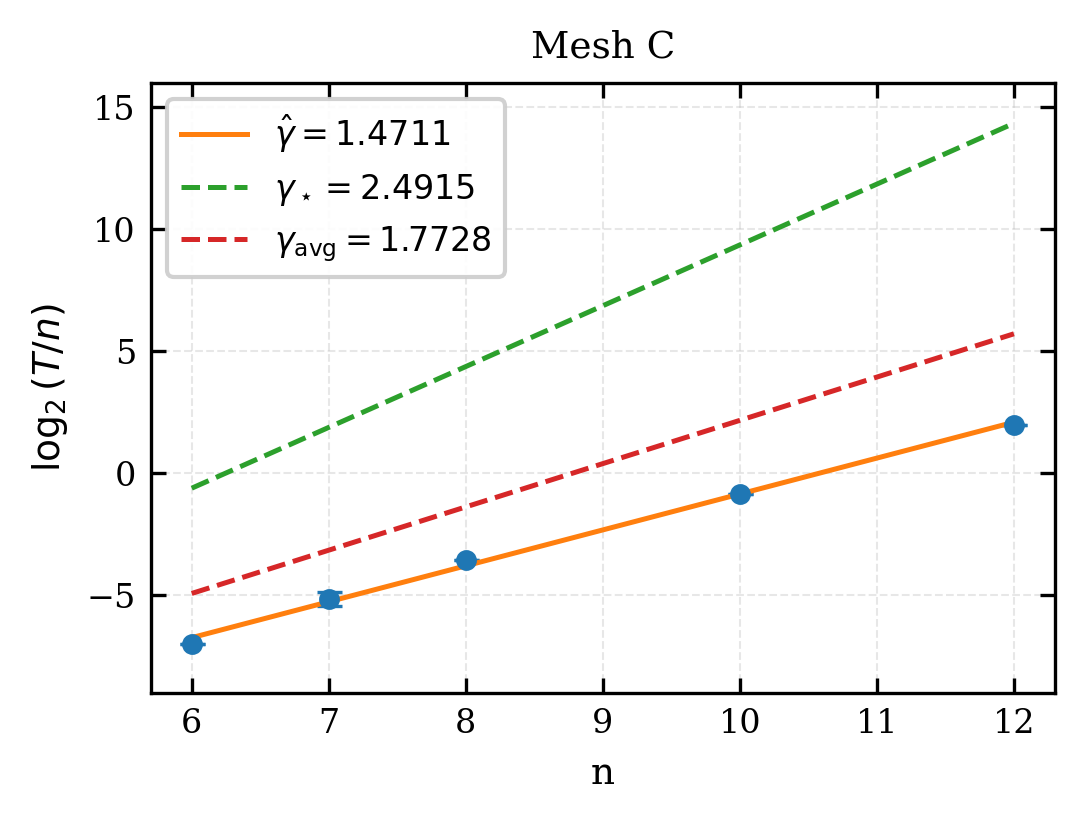}
        \caption{Mesh C: halfpipe}
        \label{fig:plot3}
    \end{subfigure}
    
    \caption{PHASE empirical runtime scaling versus theoretical bounds for three FE meshes. Orange: empirical fit $\hat{\gamma}$. Green dashed: worst-case bound $\gamma_\star$. Red dashed: average-case reference $\gamma_\text{avg}$ (not a proven bound).}
    \label{fig:empirical-results}
\end{figure}

Figure~\ref{fig:empirical-results} reports the empirical scaling exponent $\hat{\gamma}$ for all three meshes, extracted by fitting a line to $\log_2(T/n)$ versus $n$, alongside two theoretical references for each mesh. The worst-case bound $\gamma_\star$, derived in Section~\ref{sec:asymptotic-runtime-model}, is computed via Eq.~\eqref{eq:gamma-def} using the observed $\eta_\star(\delta_\star) = -\log_2 \delta_\star$; for Mesh A, where $\nu_\star > 1/2$, this falls within Region I, while for meshes B and C, isolated severely unbalanced cuts produce $\nu_\star \leq 1/2$, placing the worst-case in Region III where $\gamma_\star(\eta_\star) = 1 / \eta_\star$. The average-case reference $\gamma_{\text{avg}}$ is not a proven bound but rather the value of Eq.~\eqref{eq:gamma-def} evaluated at $\delta_{\text{avg}}$, the mean partition imbalance across all cuts; it is included to illuminate typical behavior in practical use. In all three experiments, $\hat{\gamma}$ falls below both $\gamma_\star$ and $\gamma_{\text{avg}}$, with $\gamma_\text{avg}$ providing the tighter reference in each case, consistent with the observation that isolated bad cuts drive $\gamma_\star$ above the empirically observed scaling. We note that data collection was limited to $n \leq 12$ by available hardware; all results are therefore pre-asymptotic.

\begin{table}[h]
\centering
\caption{Partition quality metrics and theoretical versus empirical scaling exponents.}
\label{tab:partition-data}
\resizebox{\textwidth}{!}{
\begin{tabular}{|l|c|c|c|c|c|c|c|c|c|}
\hline
Mesh & $\delta_\star$ & $\eta_\star$ & Region ($\star$) & $\delta_\text{avg}$ & $\eta_\text{avg}$ & Region (avg) &$\gamma_\star$ & $\gamma_\text{avg}$ & $\hat\gamma$ \\ \hline
Mesh A & 0.7000 & 0.5146 & I & 0.5953 & 0.7483 & I & 1.9725 & 1.7508 & 1.3576 \\ \hline
Mesh B & 0.7611 & 0.3939 & III & 0.5506 & 0.8610 & I & 2.5386 & 1.7046 & 1.4360 \\ \hline
Mesh C & 0.7571 & 0.4014 & III & 0.6121 & 0.7082 & I & 2.4915 & 1.7728 & 1.4711 \\ \hline
\end{tabular}
}
\end{table}

Table~\ref{tab:partition-data} illustrates how partition quality governs the tightness of the theoretical bound. Mesh A admits consistently moderate imbalance across all cuts with $\delta_\star=0.70$, placing the worst-case bound within Region I and yielding $\gamma_\star=1.97$. In contrast, Meshes B and C are pushed into Region III in the worst-case by isolated severely imbalanced cuts: in mesh B, the geometric irregularity of the crack tip produces a single cut with $\delta=0.76$ that the circular bisector cannot resolve cleanly, while in Mesh C, the curvature of the halfpipe surface causes the spherical bisector to misalign with the intrinsic 2D mesh topology at one partition level, similarly driving $\delta_\star=0.76$. Notably, despite Mesh B having better average partition quality than Mesh A ($\delta_\text{avg}=0.551$ vs $0.596$), its empirical exponent $\hat{\gamma}$ is higher, indicating that the crack-tip outliers measurably inflate the runtime beyond what average partition quality alone would predict. These observations are consistent with the guarantees of the partitioning algorithm of~\cite{miller1998geometric}, which observes balanced cuts on average but permits isolated outliers in regions of geometric irregularity; as reflected in Table~\ref{tab:partition-data}, $\delta_\text{avg}$ remains well within Region I for all three meshes, and $\gamma_\text{avg}$ consequently provides a tighter and more practically relevant upper bound than $\gamma_\star$ in each case, as confirmed by the empirical data in Figure~\ref{fig:empirical-results}.

Mesh C demonstrates that the scaling exponent $\gamma_d$ is governed by the topological dimension of the mesh rather than the dimension of the ambient embedding space. The empirical exponent $\hat{\gamma} = 1.47$ is consistent with the balanced $d=2$ theoretical prediction $\gamma_2=5/2\approx 1.67$ as expected: the asymptotic complexity bounds derived in Section~\ref{sec:asymptotic-runtime-model} depend on cut element scaling (Section~\ref{sec:asymptotic-scaling-of-cut-elts}), which is governed by the intrinsic Hausdorff dimension of the mesh regardless of how it is embedded. While the theoretical scaling exponent anchors to the intrinsic dimension, tha partition quality $\delta$, and consequently $\eta$, is sensitive to how the partitioning algorithm interacts with the mesh geometry. For Mesh C, the spherical bisector operates in the ambient $\mathbb{R}^3$ space and can align with the intrinsic 2D topology of the curved surface in ways that produce higher $\delta$ values than observed on flat 2D meshes. This explains why $\hat\gamma$ is higher for Mesh C than Mesh A despite both being topologically 2D: the difference reflects degraded partition quality on a curved surface, not a change in the underlying dimension-dependent scaling, and is captured by the correspondingly higher $\delta_\star$ and $\delta_\text{avg}$ reported in Table~\ref{tab:partition-data}. 


\section{Conclusion}
\label{sec:conclusion}

Constructing quantum-compatible representations of finite element stiffness matrices is a prerequisite for end-to-end quantum FEM pipelines, yet the exponential growth of the Pauli basis with problem size makes naive decomposition approaches computationally prohibitive for systems of engineering interest. Existing methods, including Tensorized Pauli Decomposition and Walsh-Hadamard-based schemes~\cite{hantzko2024tensorized,georges2025pauli,koska2024tree}, exploit algebraic structure but not the geometric organization in finite element discretizations, which leads to poor asymptotic scaling on FEM stiffness matrices. This work introduced PHASE, a hierarchical, geometry-aware algorithm that addresses this gap by coupling Pauli decomposition with the spatial structure of the mesh.

PHASE proceeds by applying a recursive bisection to the mesh to produce a binary partition tree whose cut-elements carry the inter-domain coupling in the stiffness operator $K$. The algorithm employs a dual-approach; it employs full-space TPD, which operates in the global $n$-qubit Hilbert space making it cost-effective at coarse depths when cut elements lack qubit overlap, and reduced-space TPD followed by FWHT-based aggregation, which exploits the qubit overlap of finer depth cut elements to reduce assembly cost. The key structural component that enables this hybrid approach is that the cut elements scale as $N^{(d-1)/d}$ at each level, a sublinear fraction of the total mesh, so the contributions from geometrically local interactions can be assembled in compressed subspaces and lifted to global Pauli coordinates via the Walsh-Hadamard transform. Balancing the two regimes at an optimal transition depth $j_\star \asymp n/(d+1)$ yields a total assembly cost $O(2^{\gamma_d n})$ with $\gamma_d = 2 - 1/(d+1)$ for balanced partitions, reducing the exponential scaling exponent relative to the $O(n 2^{2n})$ cost of applying TPD to FEM matrices generally. For two- and three- dimensional meshes this gives $O(n2^{5n/3})$ and $O(n2^{7n/4})$ respectively. This improvement is maintained provided no recursive partition produces a subdomain containing more than approximately 71\% of the nodes of its parent ($\eta>1/2$); beyond this threshold the benefit of hierarchal aggregation degrades predictably, with $O(n2^{2n/\eta})$ governing the strongly unbalanced regime.

These results represent a dimension-dependent exponential improvement in the cost of Pauli operator synthesis for finite element systems, reducing the scaling exponent from $2n$ to $\gamma_d n < 2n$ and bringing large-scale quantum-compatible assembly into a more practical computational regime. Numerical experiments on three geometrically distinct meshes confirm that empirical scaling exponents $\hat{\gamma}$ fall below both the worst-case bound $\gamma_\star$ and the average-case reference $\gamma_\text{avg}$ across all problem instances, with $\gamma_\text{avg}$ providing the tighter and more practically relevant reference in each case. We note that these results are pre-asymptotic, with data collection limited to $n \leq 12$ by available hardware, and that the complexity analysis assumes quasi-uniform meshes under shape-regularity; extension to adaptive or graded meshes would require revisiting the cut element scaling estimates. Additionally, PHASE addresses the classical preprocessing step of operator synthesis and does not itself produce a quantum circuit; translating the resulting Pauli representation into a block-encoded circuit suitable for deployment in an LCU-based solver remains a downstream task.

There are several directions that could present themselves as natural extensions of this work. The most immediate is larger-scale empirical validation: the pre-asymptotic behavior of the current experiments leaves open whether the theoretical scaling exponents are achieved in the asymptotic regime, and hardware advancements that permit $n>12$ would allow a more definitive comparison between empirical and predicted exponents, particularly for imbalance-sensitive regimes. Second, connecting the Pauli representation produced by PHASE to a deployable quantum circuit is the critical downstream step for any end-to-end quantum FEM pipeline. The LCU coefficients $\{\alpha_P\}$ and Pauli strings $\{P\}$ output by PHASE are direct inputs to block-encoding constructions, and developing circuit synthesis procedures that exploit the hierarchical structure of the PHASE output may yield further reductions in circuit depth or ancilla requirements. Finally, extending PHASE to adaptive and non-quasi-uniform meshes would substantially broaden its practical applicability: the cut element scaling $|T^\times_{[k]}| \asymp 2^{\beta_d n + \beta_d^- k}$ relies on shape regularity and quasi-uniformity, and deriving analogous bounds for locally refined meshes, where element sizes vary significantly across the domain, is a practically important problem. 

\section*{CRediT authorship contribution statement}

Tillman Philo: Conceptualization, Methodology, Software, Formal analysis, 
Investigation, Data curation, Visualization, Writing – original draft, 
Writing – review \& editing. 
Caglar Oskay: Conceptualization, Supervision, Methodology, Formal analysis, Writing – review \& editing, 
Project administration, Funding acquisition.

\section*{Data availability}

Data supporting the findings of this study are available from the corresponding author upon reasonable request.

\section*{Declaration of competing interest}

The authors declare that they have no known competing financial interests 
or personal relationships that could have appeared to influence the work 
reported in this manuscript.

\section*{Acknowledgments}
The authors gratefully acknowledge the funding support from the National Science Foundation, CMMI Division, Mechanics
of Materials and Structures Program, United States of America (Award No: 2222404 and No: 2527249).


\bibliographystyle{unsrt}
\bibliography{refs}


\appendix

\section{Analytical Derivations}
\label{app:analytical-derivations}

\subsection{Derivation of Cut Element Scaling}
\label{app:derivation-of-cut-elt-scaling}

We derive the asymptotic scaling of the number of cut elements introduced at a given depth of the recursive partition hierarchy. Throughout, we work under the mesh assumptions of Definitions~\ref{def:mesh}~and~\ref{def:conformity-etc} : $\msh$ is a conforming, shape-regular, quasi-uniform mesh of a bounded Lipschitz domain $\Omega \subset \mathbb{R}^d$, with $N = |\mathcal{E}| \asymp |\mathcal{V}|$ elements and characteristic mesh size $h \asymp N^{-1/d}$.

Let $\Sigma$ be a separator in the sense of Definition~\ref{def:separator}, with finite $(d-1)$-dimensional surface area $|\Sigma| = O(1)$. An element $e \in \mathcal{E}$ is classified as a cut element if its support intersects $\Sigma$. For a quasi-uniform mesh, any such element must lie within $O(h)$ of $\Sigma$, and the $(d-1)$-dimensional cross-sectional area of an element perpendicular to $\Sigma$ scales as $h^{d-1}$. Since shape regularity ensures these intersection patches are disjoint up to mesh-quality constants, the number of cut elements is
\begin{equation}\#\{\text{cut elements}\} \asymp \frac{|\Sigma|}{h^{d-1}} \asymp h^{-(d-1)}.\end{equation}
Substituting $h \asymp N^{-1/d}$ gives $\#\{\text{cut elements}\} \asymp N^{(d-1)/d}$.

We now extend this single-separator estimate to the recursive hierarchy. At depth $k$ of the partition tree, there are $2^k$ nonempty subgraphs, each containing approximately $2^{n-k}$ nodes. Here $n = k_{\max} + \lceil \log_2 \nu_{\max} \rceil$ is the global qubit register length (Section~\ref{recursive-application-and-dof-encoding}); under balanced bisection $k_{\max} = \lceil \log_2 N \rceil$ and $\lceil \log_2 \nu_{\max} \rceil$ is a small additive constant independent of $N$, so $2^n \asymp N$ and the subgraph size $2^{n-k}$ is proportional to $N \cdot 2^{-k}$ up to the mesh-quality constants already absorbed by $\asymp$. Applying the single-separator scaling relation to each subgraph, the number of cut elements introduced by the depth-$k$ separator within a single subgraph of size $\asymp 2^{n-k}$ is
\begin{equation}\asymp (2^{n-k})^{\beta_d} = 2^{\beta_d(n-k)},\end{equation}
where $\beta_d := (d-1)/d$. Summing over all $2^k$ subgraphs at depth $k$ and using $\beta_d^- := 1 - \beta_d = 1/d$ yields
\begin{equation}|T^\times_{[k]}| \asymp 2^k \cdot 2^{\beta_d(n-k)} = 2^{\beta_d n + \beta_d^- k}.\end{equation}
For reference, we also define $\beta_d^+ := 1 + \beta_d = (2d-1)/d$, which arises in the asymptotic complexity analyses of Sections~\ref{sec:full-space-tpd}~and~\ref{sec:reduced-space-tpd-method}. Together, $\beta_d$, $\beta_d^-$, and $\beta_d^+$ are the dimensionality scaling factors introduced in Section~\ref{sec:asymptotic-scaling-of-cut-elts}.

The scaling $|T^\times_{[k]}| \asymp 2^{\beta_d n + \beta_d^- k}$ has two notable consequences. First, since $\beta_d < 1$, the total number of cut elements at any fixed depth grows sublinearly with $N$, confirming that separators intersect only a geometrically thin layer of the mesh. Second, the factor $2^{\beta_d^- k}$ reflects the mild growth in the total interface count as progressively finer subgraphs are cut at deeper levels, and this controlled growth is the geometric property that underlies the efficiency of the PHASE assembly hierarchy.

\subsection{Full-Space TPD Asymptotic Complexity}
\label{app:full-space-tpd-runtime-derivation}

We derive the per-depth asymptotic cost of the full-space TPD assembly path described in Section~\ref{sec:full-space-tpd}. The argument proceeds in two steps: bounding the per-element decomposition cost, and multiplying by the number of cut elements at depth $k$.

For a cut element $e \in T^\times_{[k]}$, the local stiffness operator $\hat{K}_e \in \mathbb{R}^{\nu_e \times \nu_e}$ is first embedded into the global $2^n$-dimensional space via the assembly matrix $L_e$, yielding $K_e = L_e \hat{K}_e L_e^\top \in \mathbb{R}^{2^n \times 2^n}$. This embedding costs $O(2^n)$ and produces a matrix with $\mathrm{nnz}(K_e) \leq \nu_e^2 = O(1)$ nonzero entries, since $\nu_e$ is bounded by a small constant independent of $n$. Applying TPD to $K_e$ therefore costs $O(n2^n)$, since the sparsity-dependent factor in the TPD complexity reduces to $O(1)$ here.

It remains to verify that the number of nonzero Pauli terms produced does not inflate the subsequent aggregation cost. For $K_e$ with support $S = \{(i,j) : (K_e)_{ij} \neq 0\}$, a Pauli coefficient $\alpha_P = \frac{1}{2^n}\mathrm{Tr}(PK_e)$ is nonzero only if there exists $(i,j) \in S$ such that $P_{ji} \neq 0$. Defining $\mathrm{Supp}(i,j) := \{P \in \mathcal{P}_n : P_{ji} \neq 0\}$, the number of nonzero Pauli terms satisfies
\begin{equation}|\mathcal{P}(K_e)| \leq \left|\bigcup_{(i,j) \in S} \mathrm{Supp}(i,j)\right| \leq |S| \cdot \max_{(i,j)}|\mathrm{Supp}(i,j)|.\end{equation}
Each $n$-qubit Pauli matrix has exactly $2^n$ nonzero entries distributed across $2^{2n}$ positions, so on average a fixed position $(i,j)$ belongs to $4^n \cdot 2^n / 2^{2n} = 2^n$ distinct Pauli supports. Since $|S| = O(1)$, we obtain $|\mathcal{P}(K_e)| = O(2^n)$. Accumulating these $p = O(2^n)$ Pauli strings of length $n$ into a hash map costs $O(pn) = O(n2^n)$, confirming that aggregation does not exceed the decomposition cost and the total per-element runtime is $O(n2^n)$.

Multiplying by the number of cut elements at depth $k$ from Appendix~\ref{app:derivation-of-cut-elt-scaling},
\begin{equation}R_\mathrm{TPD}(n,k) = O(n2^n \cdot |T^\times_{[k]}|) = O\!\left(n \cdot 2^n \cdot 2^{\beta_d n + \beta_d^- k}\right) = O\!\left(n\, 2^{\beta_d^+ n + \beta_d^- k}\right),\end{equation}
where the simplification uses $1 + \beta_d = \beta_d^+$.

\subsection{Reduced-Space TPD Asymptotic Complexity}
\label{app:reduced-space-tpd-asymptotic-complexity}

We derive the component of per-depth asymptotic complexity of the reduced-space TPD block described in Section~\ref{sec:reduced-space-tpd-method}.

At depth $k$, each cut element $e \in T^\times_{[k]}$ admits a reduced-space operator $\tilde{K}_e \in \mathbb{R}^{2^{p_k} \times 2^{p_k}}$ with $p_k = n - k$, obtained by stripping the leading $k$ coarse-address bits from each node label (Section~\ref{sec:global-aggregation-via-fwht}, Eq.~\eqref{eq:k-tilde}). Since $\nu_e$ and the local sparsity of $\hat{K}_e$ are independent of $n$, the number of nonzero entries in $\tilde{K}_e$ is $O(1)$, and TPD applied to $\tilde{K}_e$ costs $O(p_k 2^{p_k}) = O((n-k)2^{n-k})$ per element by the same argument as in Appendix~\ref{app:full-space-tpd-runtime-derivation} with $n$ replaced by $p_k$. Multiplying by $|T^\times_{[k]}| \asymp 2^{\beta_d n + \beta_d^- k}$ from Appendix~\ref{app:derivation-of-cut-elt-scaling},
\begin{equation} \label{eq:app-tpdr}
    R_\mathrm{TPD\text{-}Reduced}(n,k) = O\!\left((n-k)2^{n-k} \cdot 2^{\beta_d n + \beta_d^- k}\right) = O\!\left((n-k)\,2^{\beta_d^+ n + k(\beta_d^- - 1)}\right),
\end{equation}
where the exponent simplification uses $(n-k) + \beta_d n + \beta_d^- k = \beta_d^+ n + k(\beta_d^- - 1)$. This is Eq.~\eqref{eq:scaling-of-local-tpd} of the main text.

\subsection{Hybrid Asymptotic Complexity Model Construction}
\label{app:hybrid-asymptotic-complexity-model}

We justify the simplified total runtime model Eq.~\eqref{eq:full-asymptotic-runtime} by showing that the reduced-space TPD term in $R_\mathrm{FWHT}(n,k)$ is asymptotically negligible, and then evaluating the resulting geometric sums.

First, we establish the asymptotic dominance of the FWHT term in the complexity model. From Appendix~\ref{app:reduced-space-tpd-asymptotic-complexity} and Section~\ref{sec:global-aggregation-via-fwht}, the two components of the per-depth reduced-space cost are $R_\mathrm{TPD\text{-}Reduced}(n,k) = O((n-k)\,2^{\beta_d^+ n + k(\beta_d^- - 1)})$ and $R_{\text{FWHT}}(n,k)=O(k\,2^{2n-k})$. To identify the dominant term, consider their ratio
\begin{equation}
\rho(k) \triangleq \frac{R_\mathrm{TPD\text{-}Reduced}(n,k)}{R_{\text{FWHT}}(n,k)} = \frac{n-k}{k}\,2^ {\Delta}, \qquad \Delta \triangleq n(\beta_d^+ - 2) + k\beta_d^-.
\end{equation}
Substituting $\beta_d^+ = (2d-1)/d$ and $\beta_d^- = 1/d$ gives
\begin{equation}
\Delta = n\!\left(\frac{2d-1}{d} - 2\right) + \frac{k}{d} = \frac{k - n}{d} \leq 0,
\end{equation}
for all $k \leq n$. Hence $\rho(k) \to 0$ as $n \to \infty$ for any fixed $k$, so the $R_{\mathrm{FWHT}}$ aggregation term dominates $R_\mathrm{TPD\text{-}Reduced}$ asymptotically. We therefore drop $R_\mathrm{TPD\text{-}Reduced}$ from the model, and the per-depth cost for $k \geq j$ reduces to $O(k\,2^{2n-k})$.

Now we examine the total cost of the hybrid complexity model. Retaining only the dominant terms at each depth, the total runtime over the hybrid scheme is
\begin{equation}
    W(n,j) \asymp \sum_{k=0}^{j-1} n\,2^{\beta_d^+ n + \beta_d^- k} + \sum_{k=j}^{n} k\,2^{2n-k},
\end{equation}
for transition depth $j$. Both sums are evaluated by identifying their dominant terms. In the first sum, we can evaluate the geometric series exactly:
\begin{equation}
\sum_{k=0}^{j-1} 2^{\beta_d^- k} = \frac{2^{\beta_d^- j} - 1}{2^{\beta_d^-} - 1} = O(2^{\beta_d^- j}),
\end{equation}
giving $\sum_{k=0}^{j-1} n\,2^{\beta_d^+ n + \beta_d^- k} = O(n\,2^{\beta_d^+ n + \beta_d^- j})$. In the second sum, the summand $k\,2^{2n-k}$ is decreasing in $k$ for $k \geq 1$, so it is dominated by its first term at $k = j$, giving $\sum_{k=j}^{n} k\,2^{2n-k} = O(j\,2^{2n-j})$. The simplified total runtime is therefore
\begin{equation}
W(n,j) = n\,2^{\beta_d^+ n + \beta_d^- j} + j\,2^{2n-j},
\end{equation}
which is Eq.~\eqref{eq:full-asymptotic-runtime}.

\subsection{Unbalanced Partition Analysis}
\label{app:unbalanced-partition-analysis}

\subsubsection{Setup}
 At each bisection of a subgraph of size $M$ nodes, let $(M_L, M_S)$ denote the child sizes with $M_L \geq M_S$ and $M_L + M_S = M$. Define the imbalance ratio
\begin{equation}
\delta := \frac{M_L}{M} \in [1/2, 1),
\end{equation}
and the log shrink factor
\begin{equation}
\eta(\delta) := -\log_2 \delta \in (0, 1], \qquad \delta = 2^{-\eta}.
\end{equation}
We assume a global imbalance cap $\delta_\star \in [1/2, 1)$ such that every split satisfies $M_L/M \leq \delta_\star$, and write $\eta_\star := -\log_2 \delta_\star$. Balanced bisection corresponds to $\delta_\star = 1/2$, i.e. $\eta_\star = 1$.

\subsubsection{Depth under imbalanced splits}

In the worst case every split achieves $\delta_\star$, so the largest subgraph after $k$ levels has size at most $\lceil \delta_\star^k N \rceil = \lceil 2^{n - \eta_\star k} \rceil$. The recursion terminates when the subgraph size falls below the prescribed threshold (Section~\ref{recursive-application-and-dof-encoding}), but it must terminate when $\delta_\star^k N \leq 1$, giving
\begin{equation}
k_{\max} = \left\lceil \frac{n}{\eta_\star} \right\rceil.
\end{equation}
This recovers $k_{\max} = n$ for balanced splits ($\eta_\star = 1$) and diverges as $\delta_\star \uparrow 1$. Since $M_k \leq \delta_\star^k N = 2^{n - \eta_\star k}$ and $p_{\max}(k) = \log_2 M_k$, the local qubit support length of the largest subgraph at depth $k$ satisfies
\begin{equation}
p_{\max}(k) \leq \max\{0,\, n - \eta_\star k\},
\end{equation}
where the $\max$ with zero reflects that support length is non-negative.

\subsubsection{Cut elements under imbalanced splits} 
Let $M_1, \dots, M_m$ denote the nodal sizes of nonempty subgraphs at depth $k$. By the geometric scaling of Eq. \eqref{eq:no-cut-elts} from Section~\ref{sec:asymptotic-scaling-of-cut-elts} the total number of depth-$k$ cut elements satisfies, up to mesh-quality constants,
\begin{equation}
|T^\times_{[k]}| \asymp \sum_{i=1}^{m} M_i^{\beta_d}, \qquad \sum_{i=1}^{m} M_i = M \asymp 2^n.
\end{equation}
This approximation is sharpest at coarse depths where subgraphs are geometrically thick; at fine depths the boundary-to-interior node ratio grows, but the resulting error is absorbed into the mesh-quality constants implicit in $\asymp$ and does not affect the asymptotic exponents. 

We derive distribution-free upper and lower bounds on $|T^\times_{[k]}|$. For the lower bound, subadditivity of $x \mapsto x^{\beta_d}$ on $[0,\infty)$ gives $M^{\beta_d} \leq \sum_i M_i^{\beta_d}$, hence
\begin{equation}
|T^\times_{[k]}| \gtrsim M^{\beta_d} \asymp 2^{\beta_d n}.
\end{equation}
For the upper bound, concavity of $x^{\beta_d}$ gives $\sum_i M_i^{\beta_d} \leq m^{1-\beta_d} M^{\beta_d}$ by Jensen's inequality. At depth $k$, the number of nonempty subgraphs is bounded by the binary refinement cap $m \leq 2^k$ and the pigeonhole cap $m \leq M$, so $m \leq \min\{2^k, M\}$. Defining the breakpoint $k_\star := \lceil \log_2 M \rceil$ and substituting $M \asymp 2^n$ (Section~\ref{sec:asymptotic-scaling-of-cut-elts}), so that $k_\star = n + O(1)$, yields
\begin{equation}
|T^\times_{[k]}| \lesssim
\begin{cases}
2^{\beta_d n + \beta_d^- k}, & k \leq n, \\
2^n, & k > n,
\end{cases}
\end{equation}
where the $O(1)$ shift in the breakpoint affects $|T^\times_{[k]}|$ by at most a fixed multiplicative constant and does not alter the asymptotic exponents.

\subsubsection{Per-depth costs} Combining the per-element costs from Sections~\ref{sec:full-space-tpd}-\ref{sec:reduced-space-tpd-method} with the cut element bound above gives the following per-depth costs. For full-space TPD,
\begin{equation}
R_\text{TPD}(n,k) \lesssim n\, 2^n |T^\times_{[k]}| \lesssim
\begin{cases}
n\, 2^{\beta_d^+ n + \beta_d^- k}, & k \leq n, \\
n\, 2^{2n}, & k > n.
\end{cases}
\end{equation}
For reduced-space TPD, the local qubit support length is bounded by $p_{\max}(k) \leq n - \eta_\star k$, so
\begin{equation}
R_{\text{TPD-R}}(n,k) \lesssim (n - \eta_\star k)\, 2^{n - \eta_\star k} |T^\times_{[k]}| \lesssim
\begin{cases}
(n - \eta_\star k)\, 2^{\beta_d^+ n + (\beta_d^- - \eta_\star)k}, & k \leq n, \\
(n - \eta_\star k)\, 2^{2n - \eta_\star k}, & k > n.
\end{cases}
\end{equation}
For FWHT aggregation, each of the $|\mathcal{L}_k| \leq 4^{n - \eta_\star k}$ local Pauli patterns requires a length-$2^k$ FWHT at cost $\Theta(k 2^k)$, giving
\begin{equation}
R_\text{FWHT}(n,k) \lesssim k\, 2^k \cdot 4^{n - \eta_\star k} = k\, 2^{2n - (2\eta_\star - 1)k}.
\end{equation}
The $k$-slopes of the latter two costs are
\begin{equation}
\frac{\partial}{\partial k} \log_2 R_{\text{TPD-R}} \asymp \beta_d^- - \eta_\star, \qquad \frac{\partial}{\partial k} \log_2 R_\text{FWHT} \asymp 1 - 2\eta_\star,
\end{equation}
so $R_{\text{TPD-R}}$ decays with $k$ iff $\eta_\star > \beta_d^-$ and $R_\text{FWHT}$ decays with $k$ iff $\eta_\star > 1/2$.

As in Appendix~\ref{app:hybrid-asymptotic-complexity-model}, we show that $R_{\text{TPD-R}}$ is asymptotically negligible relative to $R_\text{FWHT}$. On $k \in [1, n]$ the ratio satisfies
\begin{equation}
\frac{R_\text{FWHT}(n,k)}{R_{\text{TPD-R}}(n,k)} = \frac{k}{n - \eta_\star k}\, 2^{\beta_d^- n - (\eta_\star - \beta_d)k}.
\end{equation}
Since $n - \eta_\star k \leq n$, the prefactor is at least $1/n$ for $k \geq 1$. For the exponential, two cases arise. If $\eta_\star \geq \beta_d$, the exponent is minimized at $k = n$, giving
\begin{equation}
\frac{R_\text{FWHT}(n,k)}{R_{\text{TPD-R}}(n,k)} \geq \frac{1}{n}\, 2^{(1-\eta_\star)n}.
\end{equation}
If $\eta_\star < \beta_d$, the exponent $\beta_d^- n + (\beta_d - \eta_\star)k$ is increasing in $k$ and bounded below by $\beta_d^- n$, giving
\begin{equation}
\frac{R_\text{FWHT}(n,k)}{R_{\text{TPD-R}}(n,k)} \geq \frac{1}{n}\, 2^{\beta_d^- n}.
\end{equation}
In both cases the ratio grows exponentially in $n$ for any fixed $\eta_\star < 1$, so $R_{\text{TPD-R}}$ is asymptotically negligible on $[1, n]$.

On the tail $n < k \leq n/\eta_\star$, the denominator $n - \eta_\star k$ is positive and at most $n(1 - \eta_\star)$, so
\begin{equation}
\frac{R_\text{FWHT}(n,k)}{R_{\text{TPD-R}}(n,k)} \geq \frac{1}{n(1-\eta_\star)}\, 2^{(1-\eta_\star)k} \geq \frac{1}{n(1-\eta_\star)}\, 2^{(1-\eta_\star)n},
\end{equation}
which again grows exponentially in $n$. For $k > n/\eta_\star$, $p_{\max}(k) \leq 0$ and $R_{\text{TPD-R}}$ contributes no further work. Hence $R_\text{FWHT}$ dominates throughout, and the simplified hybrid total is
\begin{equation}
W(n, j) \asymp \sum_{k=0}^{j-1} R_\text{TPD}(n,k) + \sum_{k=j}^{k_{\max}} R_\text{FWHT}(n,k).
\end{equation}

\subsubsection{Region I: mild imbalance ($\eta_\star > 1/2$)} Since $2\eta_\star - 1 > 0$, the term $R_\text{FWHT}(n,k) = k\,2^{2n-(2\eta_\star-1)k}$ is decreasing in $k$, so the FWHT sum is dominated by its first term at $k = j$, and the TPD-F sum by its last term at $k = j-1$:
\begin{equation}
W(n,j) \asymp n\,2^{\beta_d^+ n + \beta_d^- j} + j\,2^{2n-(2\eta_\star-1)j}.
\end{equation}
Balancing exponents and neglecting logarithmic prefactors gives
\begin{equation}
\beta_d^+ n + \beta_d^- j \asymp 2n - (2\eta_\star - 1)j \implies j^\star = \frac{\beta_d^-}{\beta_d^- + 2\eta_\star - 1}\,n = \frac{n}{1 + d(2\eta_\star - 1)},
\end{equation}
where the second equality uses $\beta_d^- = 1/d$. At $j^\star$ both terms share exponent $\gamma(\eta_\star, d)\,n$ where
\begin{equation}
\gamma(\eta_\star, d) = 2 - \frac{2\eta_\star - 1}{1 + d(2\eta_\star - 1)},
\end{equation}
giving $W(n) = O(n\,2^{\gamma(\eta_\star,d)\,n})$. As a consistency check, $\eta_\star = 1$ recovers the balanced result $\gamma = 2 - 1/(d+1)$ from Section~\ref{sec:optimal-transition-depth}, and $\gamma \to 2$ as $\eta_\star \downarrow 1/2$.

\subsubsection{Region II: knife-edge ($\eta_\star = 1/2$)} Here $2\eta_\star - 1 = 0$, so $R_\text{FWHT}(n,k) \asymp k\,2^{2n}$ is flat in its exponential factor and the FWHT sum no longer decays with $j$. Setting $2\eta_\star - 1 = 0$ in the balance condition of Region I gives $j^\star = n$, so the TPD-F sum runs to $k = n-1$ and the FWHT sum runs from $k = n$ to $k_{\max} = \lceil n/\eta_\star \rceil = 2n$:
\begin{equation}
\sum_{k=0}^{n-1} R_\text{TPD}(n,k) \asymp n\,2^{\beta_d^+ n + \beta_d^- n} = n\,2^{2n},
\end{equation}
\begin{equation}
\sum_{k=n}^{2n} R_\text{FWHT}(n,k) \asymp \left(\sum_{k=n}^{2n} k\right) 2^{2n} \asymp n^2\,2^{2n},
\end{equation}
where the second sum uses $2^{2n-(2\eta_\star-1)k} = 2^{2n}$ and $\sum_{k=n}^{2n} k = \Theta(n^2)$. The FWHT sum dominates, giving
\begin{equation}
W(n) = O(n^2\,2^{2n}).
\end{equation}

\subsubsection{Region III: strong imbalance ($0 < \eta_\star < 1/2$)} Here $1 - 2\eta_\star > 0$, so $R_\text{FWHT}(n,k) = k\,2^{2n-(2\eta_\star-1)k}$ is increasing in $k$. The FWHT sum is therefore dominated by its last term at $k_{\max} = \lceil n/\eta_\star \rceil$:
\begin{equation}
\sum_{k=j}^{k_{\max}} R_\text{FWHT}(n,k) \asymp k_{\max}\,2^{2n-(2\eta_\star-1)k_{\max}} \asymp \frac{n}{\eta_\star}\,2^{n/\eta_\star},
\end{equation}
where the second asymptotic uses $k_{\max} \asymp n/\eta_\star$ and $(2\eta_\star - 1)k_{\max} \asymp -(1-2\eta_\star)n/\eta_\star = n/\eta_\star - 2n$. Since this dominant term is independent of $j$, the optimal strategy is $j^\star = 0$, avoiding all TPD-F cost, giving
\begin{equation}
W(n) = O\!\left(\tfrac{n}{\eta_\star}\,2^{n/\eta_\star}\right).
\end{equation}
This strictly dominates the TPD-F scale $n\,2^{\beta_d^+ n}$ for all $d \geq 2$, since $1/\eta_\star > 2 \geq \beta_d^+$ whenever $\eta_\star < 1/2$.

\subsubsection{Summary} Combining the three regimes,
\begin{equation}
W(n) \asymp
\begin{cases}
n\,2^{\gamma(\eta_\star,d)\,n}, & \eta_\star > 1/2, \quad \gamma(\eta_\star,d) = 2 - \dfrac{2\eta_\star-1}{1+d(2\eta_\star-1)}, \\[6pt]
n^2\,2^{2n}, & \eta_\star = 1/2, \\[4pt]
\dfrac{n}{\eta_\star}\,2^{n/\eta_\star}, & 0 < \eta_\star < 1/2.
\end{cases}
\end{equation}
PHASE retains an exponential improvement over naive decomposition provided $\eta_\star > 1/2$, corresponding to $\delta_\star \lesssim 0.71$: no recursive partition may assign more than approximately $71\%$ of nodes to one child.

\subsection{Correctness of the Lifting Map $\Upsilon_{\bar{q}}^{(k)}$}
\label{app:projection-proof-of-correctness}

We prove that the lifting map $\Upsilon_{\bar{q}}^{(k)}$ defined in Section~\ref{sec:reduced-space-tpd-method} satisfies $\mathcal{P}(K_e) = \Upsilon_{\bar{q}}^{(k)}\!\left(\mathcal{P}(\tilde{K}_e)\right)$, i.e., that tensoring $\Pi_{\bar{q}}^{(k)}$ onto each tuple in $\mathcal{P}(\tilde{K}_e)$ correctly recovers the global Pauli representation of $K_e$.

\subsubsection{Construction of $\Pi_{\bar{q}}^{(k)}$} The global embedding $K_e = L_e \hat{K}_e L_e^\top$ acts nontrivially only on basis states $|b_1 \dots b_n\rangle$ whose first $k$ bits match the coarse address $\bar{q}$, i.e., states of the form $|\bar{q}\rangle \otimes |\psi\rangle$ with $|\psi\rangle \in \mathbb{C}^{2^{p_k}}$. The appropriate projector onto the $|\bar{q}\rangle$ subspace of the first $k$ qubits is the rank-1 operator
\begin{equation}
\Pi_{\bar{q}}^{(k)} = \bigotimes_{i=1}^k |\bar{q}_i\rangle\langle\bar{q}_i| = \frac{1}{2^k}\bigotimes_{i=1}^k\bigl(I + (-1)^{\bar{q}_i}Z\bigr),
\end{equation}
where the second equality uses $|0\rangle\langle 0| = (I+Z)/2$ and $|1\rangle\langle 1| = (I-Z)/2$. Distributing the tensor product over the sum and factoring the phase $(-1)^{\sum_i s_i\bar{q}_i} = (-1)^{\langle s,\bar{q}\rangle}$ recovers
\begin{equation}
\Pi_{\bar{q}}^{(k)} = \frac{1}{2^k}\sum_{s\in\{0,1\}^k} \Phi_{\bar{q}}(s)\,\Xi_k(s),
\end{equation}
which is Eq.~\eqref{eq:proj-def}. This confirms that $\Pi_{\bar{q}}^{(k)}$ is expressible as a linear combination of $I$-$Z$ Pauli strings, and therefore compatible with the Pauli tuple-set representation $\mathcal{P}(\cdot)$.

\subsubsection{Correctness of $\Upsilon_{\bar{q}}^{(k)}$} 
Since $K_e$ acts as $\tilde{K}_e$ on the local $p_k$-qubit subspace and as the identity on states outside $|\bar{q}\rangle$, we have the operator identity
\begin{equation}
K_e = \Pi_{\bar{q}}^{(k)} \otimes \tilde{K}_e.
\end{equation}
Substituting the Pauli expansion $\tilde{K}_e = \sum_{(\lambda_e^\Lambda, \Lambda)\,\in\, \mathcal{P}(\tilde{K}_e)} \lambda_e^\Lambda\,\Lambda$ and using linearity of the tensor product gives
\begin{equation}
K_e = \sum_{(\lambda_e^\Lambda,\Lambda)\,\in\,\mathcal{P}(\tilde{K}_e)} \lambda_e^\Lambda\,\Pi_{\bar{q}}^{(k)} \otimes \Lambda.
\end{equation}
Expanding $\Pi_{\bar{q}}^{(k)}$ via Eq.~\ref{eq:proj-def},
\begin{equation}
K_e = \sum_{(\lambda_e^\Lambda,\Lambda)\,\in\,\mathcal{P}(\tilde{K}_e)}\, \sum_{s\in\{0,1\}^k} \frac{\Phi_{\bar{q}}(s)}{2^k}\,\lambda_e^\Lambda\,\Xi_k(s)\otimes\Lambda.
\end{equation}
Each term $\Xi_k(s)\otimes\Lambda$ is a tensor product of Pauli operators on all $n$ qubits, so this is precisely a global Pauli expansion of $K_e$. Reading off the tuple-set representation,
\begin{equation}
\mathcal{P}(K_e) = \left\{\left(\frac{\Phi_{\bar{q}}(s)}{2^k}\lambda_e^\Lambda,\, \Xi_k(s)\otimes\Lambda\right) \;\middle|\; (\lambda_e^\Lambda, \Lambda)\in\mathcal{P}(\tilde{K}_e),\; s\in\{0,1\}^k\right\} = \Upsilon_{\bar{q}}^{(k)}\!\left(\mathcal{P}(\tilde{K}_e)\right),
\end{equation}
which is the definition of $\Upsilon_{\bar{q}}^{(k)}$ in Section~\ref{sec:reduced-space-tpd-method}.

\section{Supplementary Concepts}
\label{app:supplementary-concepts}

\subsection{Geometric Implementation Notes}
\label{sec:geo-implementation-notes}
\begin{remark}[Equivalence to $\varepsilon$-displacement]
    Let $v \in \nodes$ such that $s(v)=0$ ($v$ lies on $\Sigma$) and define 
    $v_{\varepsilon} := v + \varepsilon n_{\Sigma}(v)$ for $\varepsilon>0$. Since $s$
    is differentiable with $\nabla s(v) \neq 0$ a.e. on $\Sigma$, we can use a Taylor
    expansion of $s(v)$ about $v_\varepsilon$:
    \begin{equation}
    s(v_\varepsilon) = s(v) + \varepsilon\,\nabla s(v)\cdot n_\Sigma(v) + o(\varepsilon)
    = \varepsilon \|\nabla s(v)\| + o(\varepsilon) > 0
    \end{equation}
    for all sufficiently small $\varepsilon$. Hence the convention $(\star)$
    (assigning separator nodes to $+$) is equivalent to geometrically ``pushing the node
    by $\varepsilon$ along the normal". Using a fixed alternative sign convention would
    simply swap the roles of $+$ and $-$ without affecting the algorithm.
\end{remark}

\subsection{FWHT Algorithm}
\label{sec:fwht-algo}

\begin{algorithm}
\caption{Fast Walsh-Hadamard Transform}
\begin{algorithmic}[1]
\Require Array $f[0 \dots D-1]$, where $D = 2^d$
\For{$i \gets 0$ to $d-1$}
    \State $\text{step} \gets 2^i$
    \For{$\text{mask} \gets 0$ to $D-1$ by $2 \cdot \text{step}$}
        \For{$j \gets 0$ to $\text{step}-1$}
            \State $u \gets f[\text{mask} + j]$
            \State $v \gets f[\text{mask} + j + \text{step}]$
            \State $f[\text{mask} + j] \gets u + v$
            \State $f[\text{mask} + j + \text{step}] \gets u - v$
        \EndFor
    \EndFor
\EndFor
\label{alg:fwht}
\end{algorithmic}
\end{algorithm}

Algorithm~\ref{alg:fwht} computes the (unnormalized) Walsh-Hadamard transform of a vector $f \in \mathbb{R}^D$, where $D=2^d$ and indices are identified with binary strings in $\{0,1\}^d$. The transform maps $f$ to a vector $H^{(d)}f$ defined by 
\begin{equation}
(H^{(k)}f)(s) = \sum_{t\in \{0,1\}^k} (-1)^{\langle s, t \rangle} f(t),
\end{equation}
where $\langle s,t \rangle$ denoted the bitwise inner product modulo $2$.

The algorithm proceeds iteratively over the $d$ bit positions. At stage $i$, entries whose binary indices differ only in the $i$-th bit are paired and replaced by their sum and difference. Each such operations corresponds to applying the $2 \times 2$ Hadamard matrix
\begin{equation}
H = 
\begin{pmatrix}
    1 & 1 \\
    1 & -1
\end{pmatrix}
\end{equation}
to the corresponding coordinate pairs. The nested loop structure ensures that all $2^{d-1}$ such pairs are processed at each stage without overlap.

This butterfly structure realizes the Kronecker factorization
\begin{equation}
H^{(d)} = H^{\otimes d}
\end{equation}
where $H$ is the $2 \times 2$ Hadamard matrix, and allows the transform to be computed in place using $\Theta(d2^d)$ arithmetic operations. No auxiliary storage beyond the input array is required.

The implementation shown computes the unnormalized transform. Normalized variants differ only by a global scaling factor $2^{-d/2}$ or $2^{-d}$, depending on the convention; such scaling is immaterial for the aggregation operations used in this work and is therefore omitted.


\end{document}